\documentclass[superscriptaddress,pra,aps,twocolumn,10pt]{revtex4-1}
\usepackage{bm}
\usepackage{amsmath}
\usepackage{amsfonts}
\usepackage{graphicx}
\usepackage{natbib}
\usepackage{epstopdf}
\usepackage[pdftex]{color}
\usepackage{hyperref}
\usepackage{verbatim}
\usepackage{float}
\hypersetup{
    colorlinks=true,       
    linkcolor=cyan,          
    citecolor=magenta,        
    filecolor=magenta,      
    urlcolor=cyan,           
    runcolor=cyan
}

\newcommand{\beq}{\begin{eqnarray}}
\newcommand{\eeq}{\end{eqnarray}}
\newcommand{\bes} {\begin{subequations}}
\newcommand{\ees} {\end{subequations}}

\newcommand{\ignore}[1]{}

\begin{document}

\title{Thermalization, Freeze-out and Noise: Deciphering Experimental Quantum Annealers} 

\author{Jeffrey Marshall}
\affiliation{Department of Physics and Astronomy, and Center for Quantum Information Science \& Technology, University of Southern California, Los Angeles, California 90089, USA}

\author{Eleanor G. Rieffel}
\affiliation{QuAIL, NASA Ames Research Center, Moffett Field, California 94035, USA}

\author{Itay Hen}
\affiliation{Department of Physics and Astronomy, and Center for Quantum Information Science \& Technology, University of Southern California, Los Angeles, California 90089, USA}
\affiliation{Information Sciences Institute, University of Southern California, Marina del Rey, California 90292, USA}

\begin{abstract}
By contrasting the performance of two quantum annealers
operating at different temperatures, we address 
recent questions related to the role of temperature in these
devices and their function as `Boltzmann samplers'. 
Using a method to reliably calculate the degeneracies of the energy
levels of large-scale spin-glass instances, we are able to estimate the 
instance-dependent effective temperature from the output of annealing runs. 
Our results corroborate the `freeze-out' picture which posits
two regimes, one in which the final state corresponds to a Boltzmann
distribution of the final Hamiltonian with a well-defined `effective
temperature' determined at a freeze-out point late in the anneal, and 
another regime in which such a distribution is not necessarily expected.
We find that the output distributions of the annealers do not in 
general correspond to 
a classical Boltzmann distribution for the final Hamiltonian.
We also find that the effective temperatures at different programming
cycles fluctuate greatly, with the effect worsening with problem size.
We discuss the implications of our results for the design of future
quantum annealers to act as more effective Boltzmann samplers 
and for the programming of such annealers.
\end{abstract}
\maketitle

\section{Introduction}
A handful of recent studies suggest that quantum annealers may
be well suited to function as fast
thermal samplers~\cite{Amin:2015,Amin:boltzmann,perdomo,fairInUnfair}.
By taking advantage of their finite temperature nature 
~\cite{katzgraber:seekingSpeedup,scirep15:Martin-Mayor_Hen,perdomo,marshall:16,fairInUnfair},
potentially they may sample from Boltzmann distributions
of certain cost functions more efficiently 
than can be done classically. 
Such a capability opens up the exciting possibility of applications of
quantum annealing to so-far-uncharted avenues of research, with immediate
applications to domains such as deep learning networks and restricted Boltzmann machines
~\cite{perdomo,adachi,Amin:boltzmann}. 

The main mechanisms that determine the distributions from which output 
configurations are drawn are thus far unclear. 
Further insights into the role of temperature, and the capabilities of 
experimental quantum annealing optimizers to quickly thermalize, are
challenging to obtain due to the limited ability to probe the inner workings 
of these machines, as well as the lack of control over
most operating parameters~\cite{perdomo,adachi,fairInUnfair}. 
 
To circumvent these difficulties, we devised an experiment,
directly comparing the performance of \emph{two} commercially available 
quantum annealers operating at different temperatures (we shall refer to those as `hot' and `cold' henceforth).
This key difference, 
together with a newly devised method
to accurately calculate the degeneracies of certain
large-scale spin-glass instances,
offers us a unique opportunity to study the effects of temperature.
Our results indicate that 
these instances do not in general equilibrate at
Boltzmann distributions corresponding to the final 
classical Hamiltonian,
but are significantly affected by nonzero quantum fluctuations and noise.  
Our results corroborate the 
`freeze-out' picture~\cite{johnson:11,Amin:2015,Amin:boltzmann},
which posits one regime in which the final state corresponds to a
Boltzmann distribution of the final Hamiltonian with well-defined
`effective (classical) temperature' determined at a generally unknown
freeze-out point late in the anneal, and another regime in which such a
distribution would not necessarily be expected.
While providing evidence for this picture, our results speak against the hypothesis
that most instances fall in the first regime.

We find that these effective temperatures fluctuate greatly at different
programming cycles, with the effect worsening with problem size. We discuss
factors potentially contributing to this
adverse effect, including so-called $J$-chaos in which control errors and other sources of noise mean
that the problem run on the machine is different from the one programmed in.
We discuss the implications of our results for the design of future
quantum annealers to act as efficient Boltzmann samplers and for the 
programming of such annealers.
\\
\subsection{Quantum annealing and quantum annealers}
\begin{figure}[htp]
\begin{center}
\includegraphics[width=.98\columnwidth]{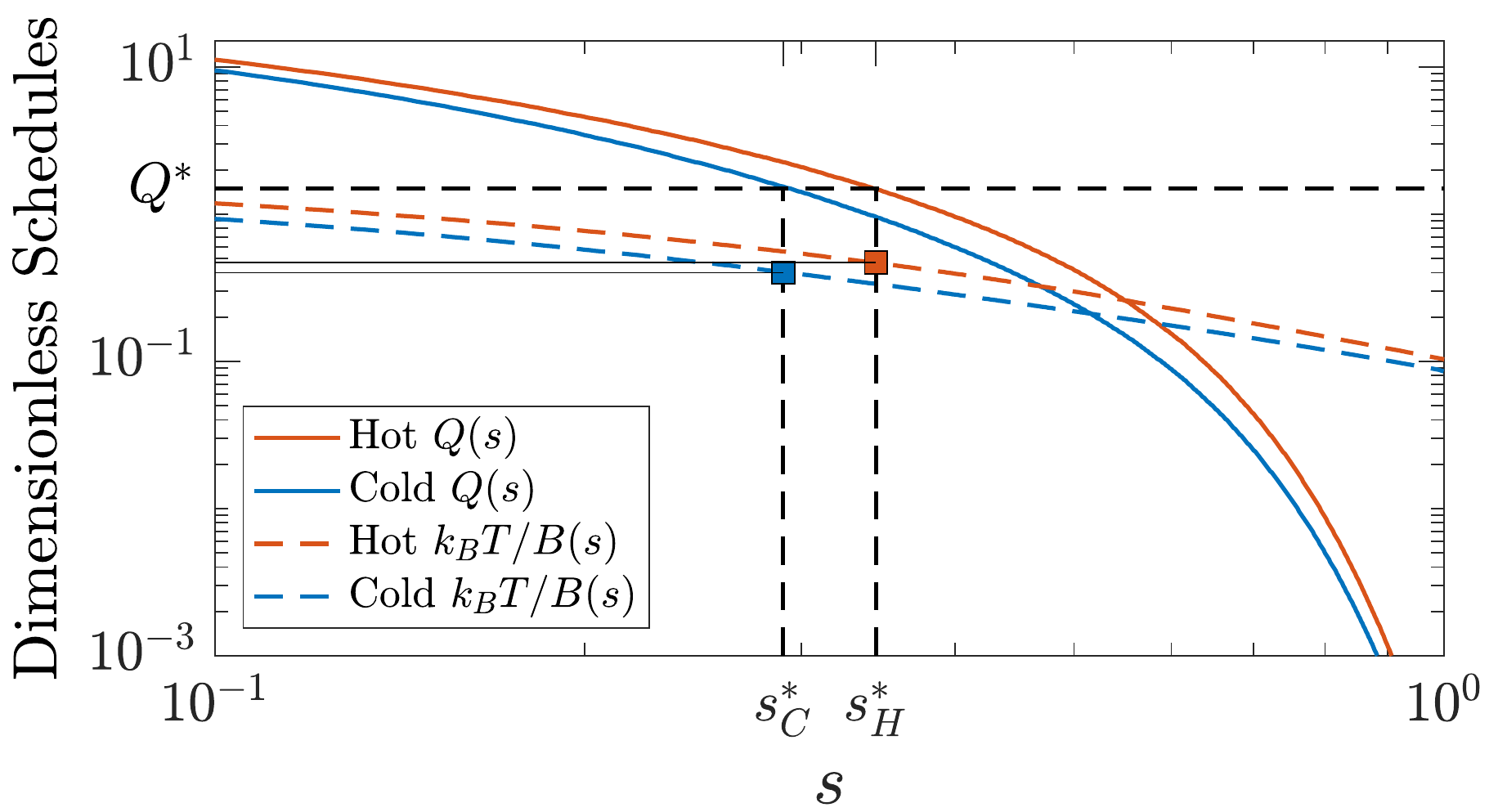}
\caption{\textbf{Dimensionless annealing schedules and temperatures of the hot and cold 
DW2 processors.} 
We plot quantum fluctuations $Q(s)$, and thermal fluctuations $k_BT/B(s)$, as a function of rescaled annealing time, $s$, for each machine.
An example of freeze-out is
shown; the distribution approximately halts at some fixed, instance-dependent
value of $Q^*:=Q(s^*)$ (dashed black line) which corresponds to a 
freeze-out point $s^*$ (denoted $s_C^*,s_H^*$) and dimensionless effective
temperature $k_B T/B(s^*)$ (squares), for each machine. 
The solid black lines illustrate that freeze-out may correspond to 
different effective temperatures. 
}
\label{fig:schedule}
\end{center}
\end{figure}

Standard transverse
field quantum annealing works by evolving the
system over rescaled time $s=t/\mathcal{T} \in[0,1]$ where $t$ is time and
$\mathcal{T}$ is the overall runtime of the annealing process. The total
Hamiltonian of the system is given by 
\beq
H(s)=A(s)H_d+B(s)H_p \,,
\label{annealing}
 \eeq
where 
$H_p=\sum_{\langle i,j\rangle} J_{ij} \sigma_i^z \sigma_j^z + \sum_i h_i \sigma_i^z$ 
is the programmable Ising spin-glass problem (the final Hamiltonian) to be sampled defined
by the parameters $\{ J_{ij},h_i\}$, and $H_d=-\sum_i \sigma^x_i$ is a
transverse-field Hamiltonian which provides the quantum fluctuations (the initial Hamiltonian). We
identify two dimensionless scales associated with the annealing, namely, the
one associated with quantum fluctuations $Q(s)=A(s)/B(s)$ and the scale
associated with thermal fluctuations $k_B T/B(s)$, both of which are shown
in Fig.~\ref{fig:schedule} for both the `hot' and `cold' processors.

Current quantum annealers suffer from intrinsic control errors
(ICE)~\cite{King:2014uq,scirep15:Martin-Mayor_Hen} such as imperfect digital-to-analog conversion
when programming the problem parameters onto the machine, and $1/f$-noise whose effect is parameter changes even within a single programming cycle (a consecutive batch of anneals run on the machine)~\cite{oneOverF1,oneOverF2}. For both contrasted quantum annealers, 
these random errors may be approximated as normally distributed
according to $\sim \mathcal{N}(0,0.05J)$ [resp. $\sim \mathcal{N}(0,0.03h)$]
where $J$ (resp. $h$) is the maximal value over
all the programmed $J_{ij}$ (resp. $h_i$). 
Some problems have resilience to such errors 
\cite{katzgraber:seekingSpeedup,venturelli2015}, whereas others are 
susceptible to a phenomenon referred to as $J$-chaos, in which 
output `solutions' correspond to the wrong, or malformed,
problem, generally reducing the success probability~\cite{nifle:92,ney-nifle:98,krzakala:05,katzgraber:07,venturelli2015,scirep15:Martin-Mayor_Hen}.

\subsection{Freeze-out conjecture}
If problems thermalized instantly,
quantum annealers would return configurations sampled
from a Boltzmann distribution, in which each configuration $c$ has weight 
proportional to $e^{-\beta^{\text{eff}}_{\text{ideal}}E_c}$, where $E_c$ is the configuration's classical cost (under $H_p$) and   
$\beta^{\text{eff}}_{\text{ideal}} \equiv B(1)/k_B T$ is
an effective dimensionless inverse temperature,
with $T$ being the operating temperature of the machine~\cite{tempScalingLaw}. 
It is known, however, that effective inverse-temperatures $\beta^{\text{eff}}$ 
extracted from experimentally sampled distributions are usually
lower than $\beta^{\text{eff}}_{\text{ideal}}$, and that the
observed inverse-temperatures differ across problem
instances~\cite{Amin:2015,Amin:boltzmann,perdomo}.  

The freeze-out conjecture \cite{johnson:11,Amin:2015,Amin:boltzmann}
explains these high observed effective temperatures   
by positing a ``small $Q$ regime'' in which the evolution 
is quasi-static,
returning a final population that is close to a Boltzmann distribution 
of $H_p$ with a well-defined effective temperature, and a regime in which
the final population would not necessarily be of this form.
In the first regime, the final distribution is determined by a `freezing' 
of the evolution (after which no dynamics occur) 
at an unknown, instance-dependent, but physical-temperature independent, 
`freeze-out' point $s^*$ where thermal fluctuations, whose strength is 
coupled to the
quantum fluctuations $Q(s^*)$ driving the system, become negligibly small~\footnote{The term `effective temperature' is somewhat of a misuse as it may imply thermalization of the system whereas in fact it may not be the case.}.  

As illustrated in Fig.~\ref{fig:schedule}, the freeze-out point is
conjectured to happen at a temperature-independent (but instance-dependent)
value $s^*$~\cite{Amin:2015}. Only when $Q(s^*)$ at the freeze-out is
small is the final distribution expected to be a classical Boltzmann
distribution for $H_p$ with (dimensionless) effective temperature 
$\beta^{\text{eff}}=B(s^*)/k_B T$;
otherwise, the resultant distribution will generally not
correspond to an equilibration at any given point, but may instead result
from different parts of the system equilibrating at different 
temperatures and times~\cite{Amin:2015}. 
\\
\subsection{High-level approach}
We proceed by taking as a working hypothesis that most instances have a 
well-defined freeze-out point in the range $A(s) \ll B(s)$. We work through
the implications of this hypothesis, and demonstrate empirically that
it does not hold for the majority of instances. We do so by estimating
a freeze-out point from the data for each instance, and then 
checking whether or not that point falls in the $A(s) \ll B(s)$ regime. Most
instances fail this consistency check. Outside of that regime, the 
freeze-out conjecture does not predict a well-defined freeze-out point;
different parts of the system may freeze at different times, and even if
an instance does have a well-defined freeze-out point outside $A(s) \ll B(s)$,
the distribution would have a strong quantum component (from $H_d$), so would be a 
distribution of quantum states, and not of the form 
$e^{-\beta^{\text{eff}}H_p}$. These results are consistent with the
freeze-out conjecture, but not with the hypothesis that most instances
fall within the freeze-out regime that yields a classical Boltzmann 
distribution.

\section{Experiment and methods}

We make use of two $512$-qubit D-Wave Two (DW2) quantum 
annealers~\footnote{One machine is owned
by Lockheed-Martin, housed at USC's Information
Sciences Institute and the other, by a NASA-USRA-Google collaboration
and housed inside the NASA Ames Research Center.}. The mean temperatures of the
`hot'  and `cold' machines were about 16.0 mK and 13.2 mK, respectively
(further details are provided in Appendix~\ref{sect:devices}).

\begin{figure}[htp]
\begin{center}
\includegraphics[width=\columnwidth]{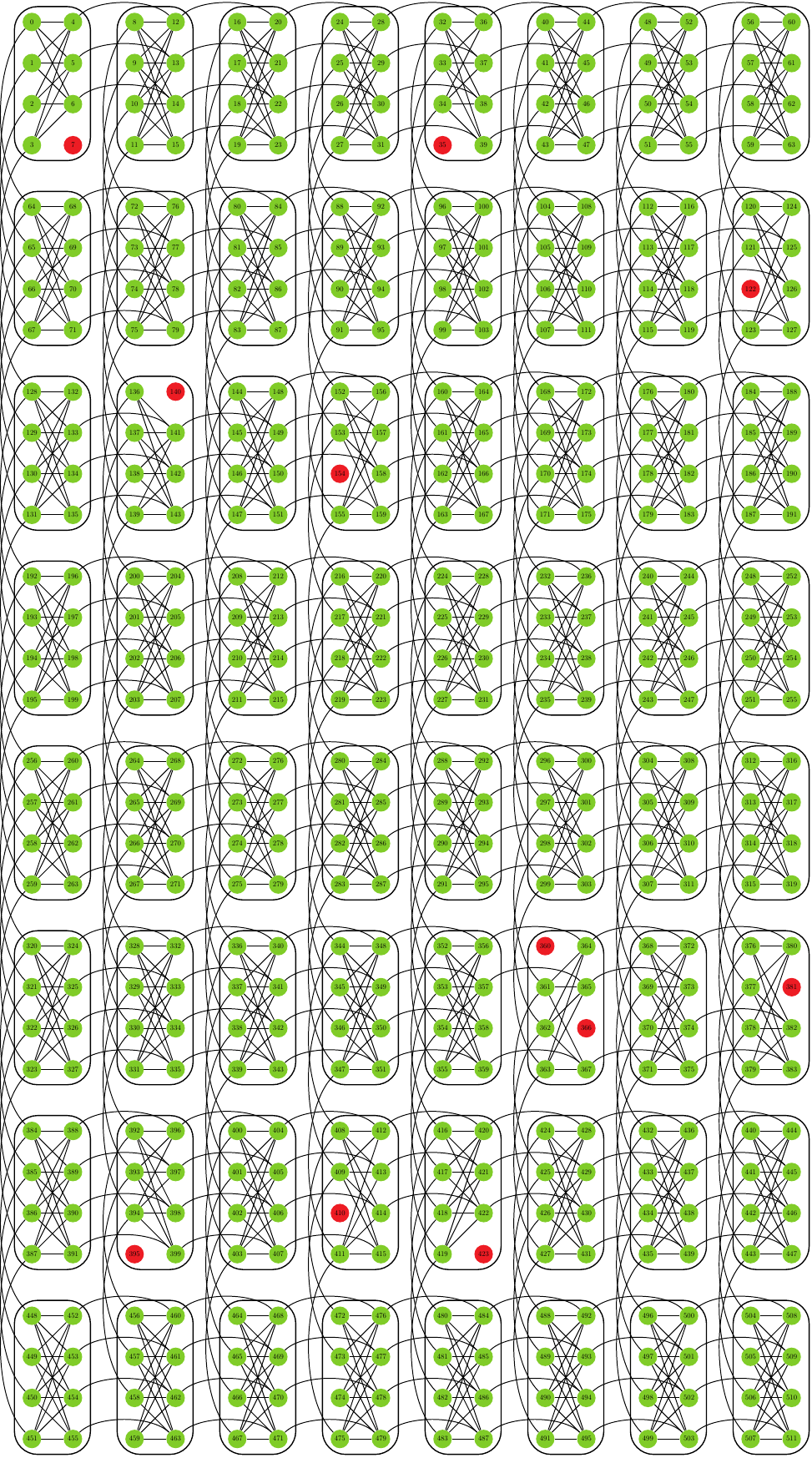}
\caption{\textbf{Intersected Chimera}. Intersection of the two D-Wave Chimera
graphs, with 501 operating qubits. Red disks denote non-operational qubits on
one (or both) of the two machines. We denote the size of a square subgraph by an integer $L \le 8$.} \label{fig:chimera}
\end{center}
\end{figure}

We designed $1300$ random spin-glass instances of the
planted-solution type~\cite{hen:15} for each of seven different problem sizes
corresponding to $L \times L$ grids of $8$-qubit cells of the hardware DW2
Chimera graph with $L=2\ldots 8$ 
(see Fig.~\ref{fig:chimera}).
We generated these instances as per Ref.~\cite{hen:15} (the reader is referred to Appendix~\ref{sec:instances} for more details).
This class of
instances is particularly suitable for our purposes for two main reasons: i)
the ground state energies of the generated problems are known in advance, and
ii) the \emph{exact} degeneracies of the ground and first excited states are
computable~\cite{fairInUnfair}.
These two facts allow us to, with high accuracy and confidence, measure $\beta^{\text{eff}}$,
as will be explained below.
We generated instances on the intersection of
the two hardware graphs ($501$ qubits) in order to
avoid biases associated with malfunctioning qubits on either machine (as shown in Fig.~\ref{fig:chimera}).

To gather our statistics, each instance was run 440,000 times over 22 `programming cycles' on each machine,
with anneal times in range [20-40] $\mu$s.
A programming cycle consists of running the same instance sequentially on a single machine up to (as chosen by the user) 20,000 times, from which statistics are returned; from each programming cycle we obtained the ground state success probability 
(how often the ground state of the problem was found). 
We use this data to estimate $\beta^{\text{eff}}$.

\begin{figure}[htp]
\begin{center}
\includegraphics[width=\columnwidth]{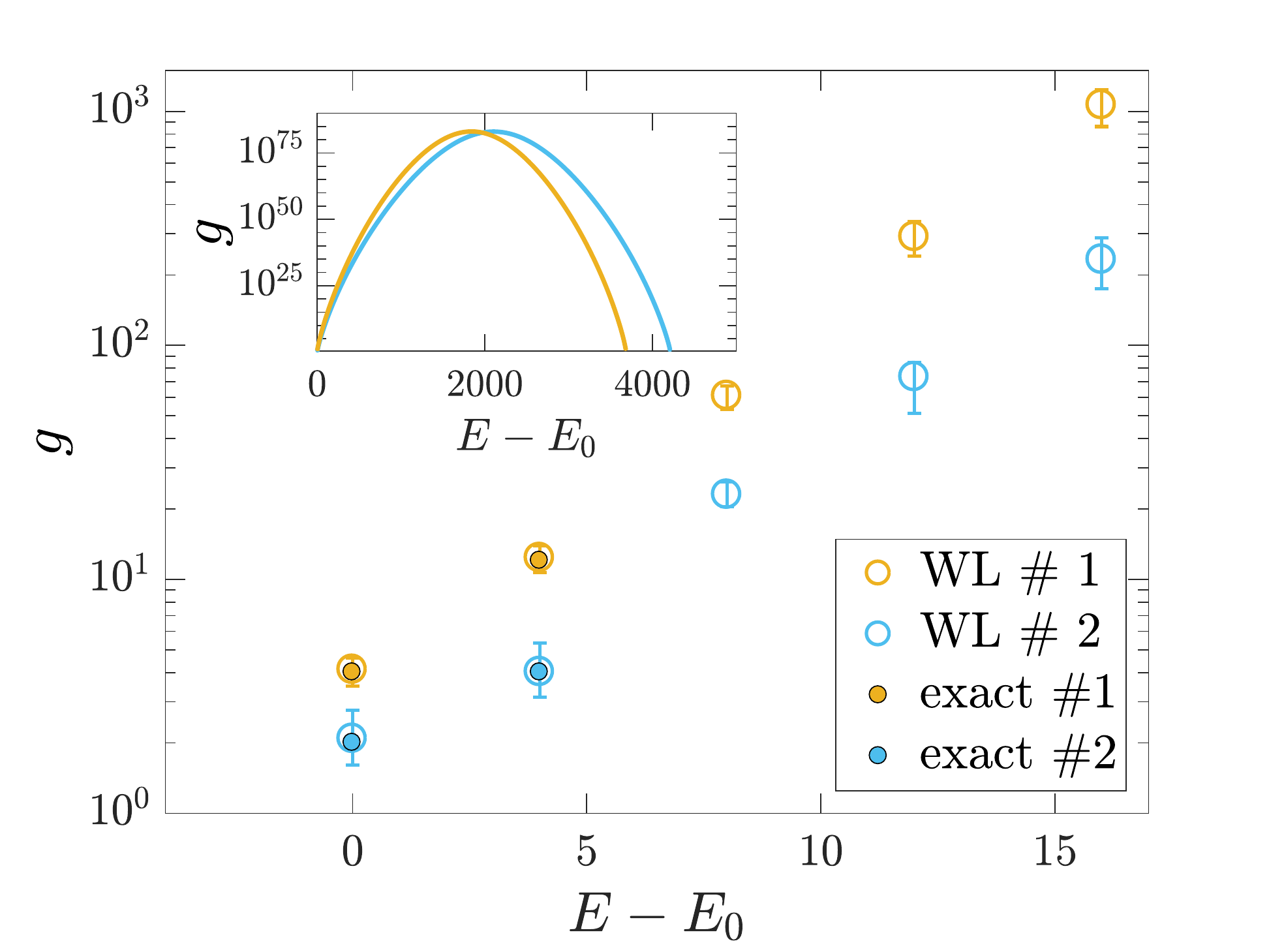}
\caption{\textbf{Degeneracy counting}. Main figure: The degeneracies of the
first five energy levels of two of the $282$-qubit problem instances as found by the Wang-Landau algorithm (error
bars represent 95\% confidence interval). The WL degeneracies of the first two
levels lie on top of the computable exact values (solid circles). Inset:
Degeneracies of all levels as a function of energy, for the same instances, as
obtained by the WL algorithm.} 
\label{fig:dcf} \end{center} 
\end{figure}

To evaluate $\beta^{\text{eff}}$
we employ two independent, complementary, techniques, which together
allow us to estimate with high accuracy and confidence the degeneracies of 
the energy levels of the problem instances. 
The first, the well-known Wang-Landau (WL) entropic
sampler~\cite{wang:01a}, statistically estimates
the degeneracy of the energy levels (see Appendix~\ref{sect:WL} for technical details). Since the WL algorithm is prone to
statistical errors as well as false convergences, we employ in parallel a 
newly-devised algorithm that 
uses the feature that planted-solution
instances can be written as a sum of local
terms~\cite{fairInUnfair}. 
The algorithm 
computes the degeneracies of
the ground and first excited states \emph{exactly}. 
When the WL estimate is outside $\pm 5\%$ of the exact value for either the
ground or first excited state, we discard this instance as we know it has not
converged properly.  
The combination of these two algorithms
allows for the faithful estimation of the degeneracies. 
We show in Fig.~\ref{fig:dcf} an example of a successful implementation of these 
two algorithms, where the Wang-Landau ground and first excited estimates are within
5\% error of the exact values. 

The above procedure yielded some 2200 instances in total, for problem sizes up to 282 qubits, 
for which we were able to accurately calculate the degeneracies.
The difficulty in obtaining an accurate measurement, especially for the larger problems, was due mainly to i) the D-Wave machine not being able to solve many of the `hard' problems, ii) 
there were too many degenerate states for the exact counter to enumerate (exceeded our chosen
cut-off value of $10^7$ ground states, which become prohibitively expensive to compute), or iii) Wang-Landau estimate deviated too far from exact counter results (generally from under-sampling the low energy states).

Armed with these degeneracies, we estimate the inverse-temperature $\beta^{\text{eff}}$ for each instance by minimizing the distance between the observed ground state success probability $P_0$
and the predicted one (i.e., the conjectured Boltzmann distribution):
\beq
\label{eq:beta}
\left| P_{0} -\left(\sum_{k=0}\frac{g_k}{g_0}e^{-\beta^{\text{eff}}(E_k-E_0)}\right)^{-1} \right| \,.
\eeq
Here, $\{g_k, E_k\}$ are the degeneracy and energy of the $k$-th
level, respectively. 
The total number of instances for which $\beta^{\text{eff}}$ was successfully estimated, for each problem size $L=2\dots 8$, is  [664,745,449,266,38,0,0].

\section{Results and Analysis}
\subsection{Thermalization}
\begin{figure}[htp]
\begin{center}
\includegraphics[width=0.85\columnwidth]{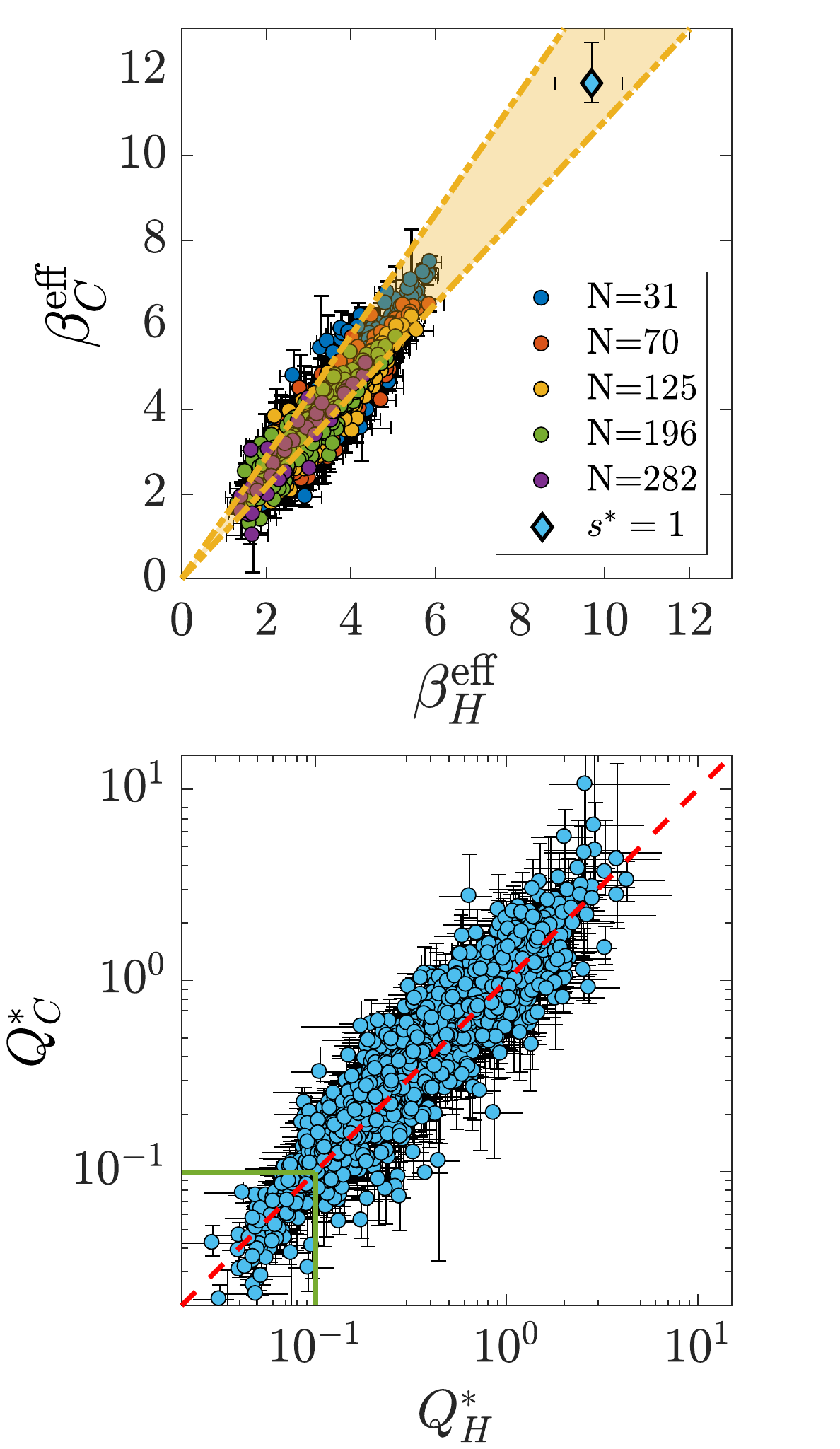}
\caption{\textbf{Top: Effective inverse temperatures}. Comparison of effective
inverse temperatures on the set of instances for both the hot processor, 
$\beta^{\text{eff}}_H$, and the cold processor, 
$\beta^{\text{eff}}_C$. Error bars represent the highest and lowest values
found over all programming cycles. The yellow band represents the range of
physical temperature fluctuations between the devices, given by the linear relationship $y=R_\beta^{\mathrm{ideal}} x$ (defined in main text). The blue diamond represents $\beta^{\text{eff}}$ 
were (classical) thermalization to take place at the end of the anneal. 
Different problem sizes have different colors (see legend).
\textbf{Bottom: Quantum fluctuations at freeze-out.} Scatter plot of the
extracted $Q^*:=Q(s^*)$ strengths (blue) for each instance. 
Ideally, both machines would yield the same value for each instance
(red $y=x$ line). Median ratio is $R^{\text{small}}_Q=1.01$
for small $Q(s^*)$ values (within the green square). The outputs are strongly
correlated (Pearson coefficient 0.92). Error bars represent the range
of $Q^*$ over all programming cycles.} \label{fig:effectiveBetas}
\end{center}
\end{figure}

Figure~\ref{fig:effectiveBetas} (top) plots the median inverse temperature
$\beta^{\text{eff}}$ for each instance and machine. Error bars indicate the
maximum and minimum value of $\beta^{\text{eff}}$ over all programming
cycles. Evident is the overall strong linear correlation between the
(inverse) effective  temperatures of the two machines
(Pearson coefficient 0.94). 
Most instances fall
within, or near, the `thermal range' (see caption) predicted by the ratio of physical
temperatures of the machines [see yellow band in 
Fig.~\ref{fig:effectiveBetas} (top)], illustrating the key functional role of temperature in the success
probability of these problems.  If the instances were
thermalizing at the end of the anneal, however, we would expect to observe
$\beta^{\text{eff}}_{\text{ideal}}$ of 9.7 and 11.7 
(shown in Fig.~\ref{fig:effectiveBetas})
for the hot and cold machines, respectively.
Instead, the values we observe are well below this mark: 
$\beta^{\text{eff}}\in[2,7]$.
Thus, we are finding effective temperatures up to six times higher
than would be expected from a simple thermalization picture.
Moreover, the median ratio of $\beta^{\text{eff}}$ for the two machines, 
$R_{\beta}=1.11 \pm 0.05$ (95\% confidence interval) \footnote{If we calculate the ratio via the (least squares) gradient of 
Fig.~\ref{fig:effectiveBetas} (top), we find it is $R=1.14$, 
also far below the physical ratio.}, 
is well below the ratio of the physical temperatures,
$R_{\beta}^{\text{ideal}}= 1.21 \pm 0.02$ indicating an effective
average temperature ratio of about $92\%$ of the `thermal' ratio of $s=1$.
We now examine the extent to which the freeze-out picture can explain these
discrepancies.

\subsection{Freeze-out}
While the freeze-out point for each instance is unknown, its temperature
independence means the estimates for the freeze out point should be the same
whether from the cold machine or hot machine data.
Using the estimated $\beta^{\text{eff}} = B(s^*)/k_BT$, from which we can
obtain a freeze-out point $s^*$ given the known operating temperatures and
annealing schedules, we directly calculate $Q(s^*)$. We then check whether
the freeze-out point for each instance is the same for the two machines.
We plot $Q(s^*)$ for each instance in Fig.~\ref{fig:effectiveBetas} (bottom). 
For instances with small $Q$ (we take $Q< 10^{-1}$), 
we find excellent correspondence between the two machines, with an average
ratio of $R^{\text{small}}_Q=1.01 \pm 0.06$ (95\% confidence interval), in
agreement with the freeze-out picture, suggesting a meaningful 
$\beta^{\text{eff}}$, which implies final classical Boltzmann distributions,
in this regime.
Only a small fraction of the instances, however, correspond to a 
negligible $Q(s^*)$.

For the majority of instances, $Q(s^*)>1$, 
and thus contradicts our working hypothesis that most instances
fall in the regime in which one would expect a well-defined freeze-out point
and a final classical Boltzmann distribution.
working assumption that the instances thermalize according to $H_p$.
The ratio $R_Q$ over the 
the entire data set is $R_Q \approx 1.20$, substantially 
higher than the `ideal' $R_Q=1$. 
Compared to the rest of the instances, the small $Q(s^*)$ problems are
typically easier to solve and are disproportionately smaller in problem size
(see e.g. Fig.~\ref{fig:Qnq} in Appendix~\ref{sect:additional}).
The freeze-out picture can also explain the
lower-than-ideal effective inverse-temperature ratio $R_{\beta}=1.11$ 
(and 
higher $R_Q \approx 1.20$). 
The existence of significant quantum fluctuations outside $A(s) \ll B(s)$
leads to an overestimation of thermal fluctuations
in both machines, i.e., to higher effective temperatures, 
as we indeed observe.

\subsection{High variability in inverse temperature estimates}
The magnitude of the error bars on the effective inverse temperatures per
instance shown in Fig.~\ref{fig:effectiveBetas} (top) reflect the large
fluctuations in success probabilities between programming cycles. 
We discuss various factors that contribute to that variance.

\begin{figure}[htp]
\begin{center}
\includegraphics[width=0.85\columnwidth]{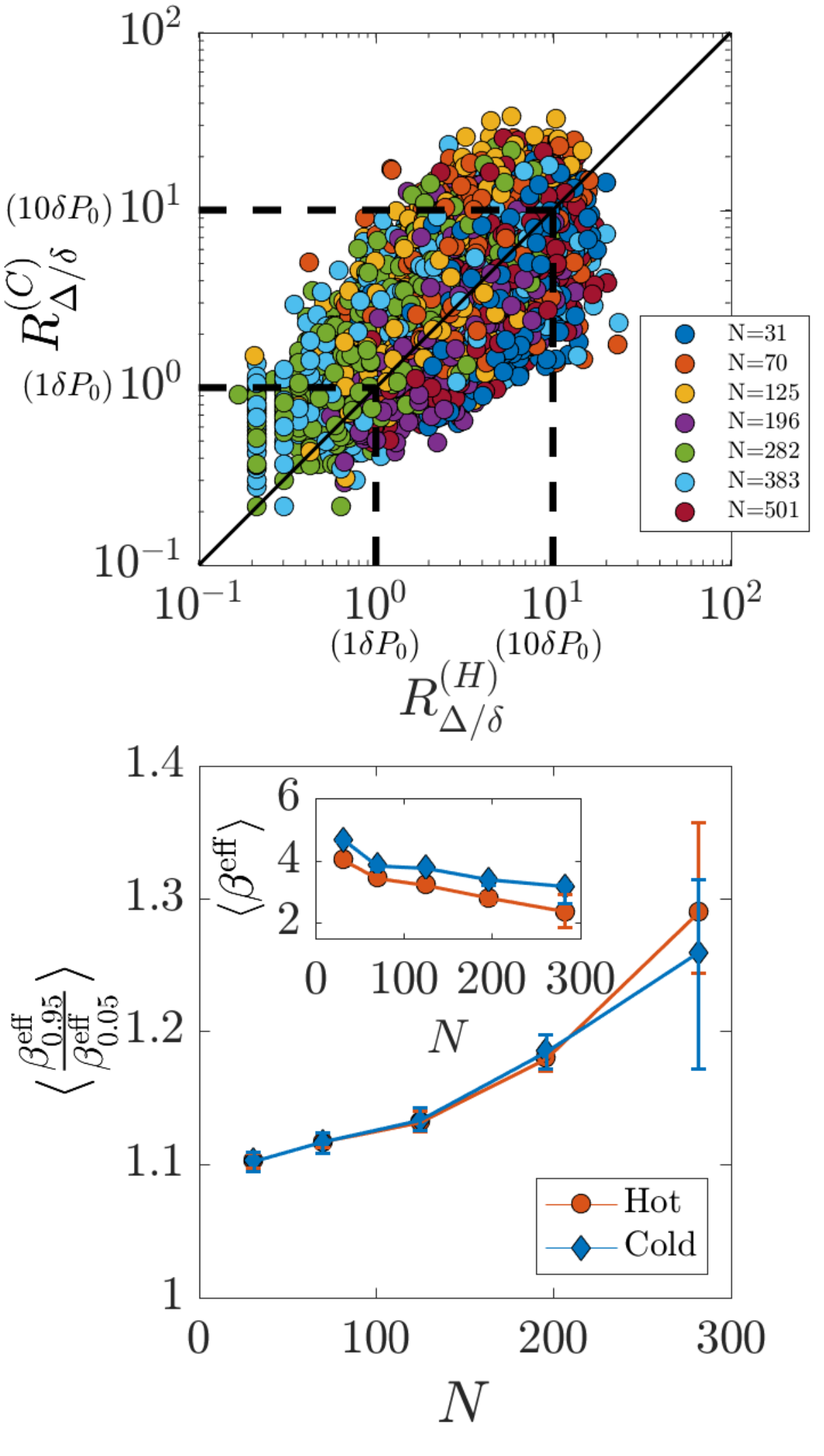}
\caption{\textbf{Top: Ratio of magnitude of actual fluctuations in success probabilities to magnitude of statistical fluctuations, $R_{\Delta/\delta}$, for the various instances on the hot and cold processors.} For most instances, the fluctuations in success probabilities over programming cycles is far greater (by an order of magnitude) than what one might expect from fluctuations of 
a purely statistical nature.   
\textbf{Bottom: Typical spread of effective inverse temperature as a function
of problem size}. Our measure for spread is the $95$th to $5$th 
percentile mean ratio of
$\beta^{\text{eff}}$ averaged over instances of each problem size.  
We take the ratio to overcome any bias from the cold chip recording
higher values of $\beta^{\text{eff}}$ (see inset). We find both
devices follow a nearly identical trend: fluctuations increase and 
$\beta^{\text{eff}}$ decreases with problem size.
Inset: Median $\beta^{\text{eff}}$ for each problem size.
95\% confidence interval error bars obtained by bootstrapping over the
instances of each particular $N$.} 
\label{fig:betaRangeRatio}
\end{center}
\end{figure}

It is known that the location of the freeze-out point (and hence the success probability) has a weak
logarithmic dependence on the annealing time
\cite{Amin:2015,scirep15:Martin-Mayor_Hen}, with 
longer anneal times having later freeze-out points because there is
more time for fluctuations to take place. We indeed find such an effect (see
Figs.~\ref{fig:annealTime}, \ref{fig:annealTime70} of Appendix~\ref{sect:additional}), though our results show that this typically accounts
for less than a 1$\%$ variability between different anneal times and therefore
does not explain the spread we observe.  If the variation were due
purely to statistical variations from cycle to cycle, one
would expect statistical fluctuations in success probability $P_{0}$ on the
order of $\delta P_0 =  \sqrt{P_0(1-P_0)/N_{\text{anneals}}}$.
Fig.~\ref{fig:betaRangeRatio} (top) shows $R_{\Delta/\delta}=\Delta P_0/\delta
P_0$, the ratio of typical magnitude of actual fluctuations in success
probabilities $\Delta P_0$ to the expected magnitude of purely statistical
fluctuations $\delta P_0$. We find that only around 20$\%$ of the instances
exhibit fluctuations of success probability $R_{\Delta/\delta}$ below 1. For
most instances, typical fluctuations are about an order of magnitude
greater than statistical fluctuations, with some fluctuations being considerably
greater. We attribute these large ratios, to
$J$-chaos~\cite{scirep15:Martin-Mayor_Hen} from ICE and other noise, which 
affect the local fields and coupling parameters within and between cycles.
Noise unrelated to programming parameters may also play a role.

\begin{figure}[htp]
\begin{center}
\includegraphics[width=0.9\columnwidth]{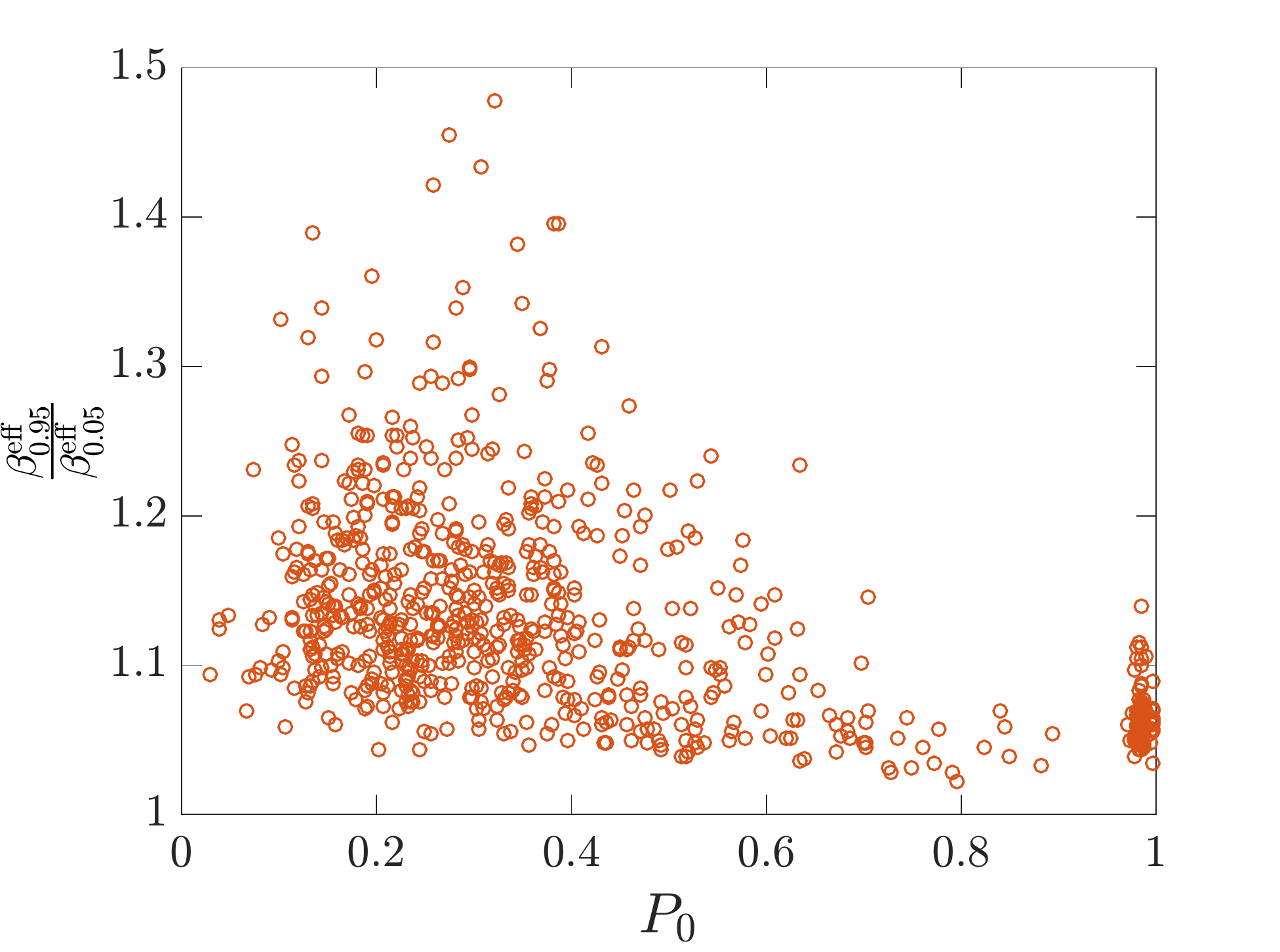}
\caption{\textbf{Variance with success probability}. Spread of effective inverse temperature as a function of success probability, as determined by the $95$th to 5th percentile ratio of $\beta^{\text{eff}}$ over all programming cycles, for $N=70$ (we pick this problem size as it is the one for which we have the most number of instances with reliable data). Data from the hot machine.}
\label{fig:varP0}
\end{center}
\end{figure}

Figure~\ref{fig:betaRangeRatio} (bottom) shows,
as a function of problem size, the average variation in $\beta^{\text{eff}}$,
as measured by the ratio of the 95th to 5th percentile values found over all
programming cycles.  
The larger the problem size, the
greater the size of the fluctuations.
This trend is expected as larger problems, with more couplings, have more
potential to be adversely affected by control errors, and other sources
of noise~\footnote{The increase in
fluctuations with problem size we observe in 
Fig.~\ref{fig:betaRangeRatio} (bottom) is most likely an underestimate of the
full effect. 
Since our criterion for discarding instances is convergence of the
WL algorithm, those instances that
do not appear in the figure exhibit fluctuations of larger magnitudes, as there
is a known strong positive correlation between WL convergence, i.e., its
classical hardness, and
$J$-chaos (see, e.g., Ref.~\cite{scirep15:Martin-Mayor_Hen}).}. 
It is critical to understand why these fluctuations scale
with problem size, and their root cause, so as to devise strategies
to keep these errors from becoming unmanageable as chip sizes increase.
For a fixed problem size, we do not observe a clear correlation
between success probability and the variance in the $\beta^{\text{eff}}$
estimates (Fig.~\ref{fig:varP0}), providing evidence that the fluctuations we
observe in Fig.~\ref{fig:betaRangeRatio} (bottom) are indeed due to differences
in problem size and not problem difficulty (though of course the two 
are related)~\footnote{We also discount other minor effects, such as 
the known logarithmic dependence on anneal time, in Appendix \ref{sect:additional}.}.

\section{Conclusions}
By conducting parallel experiments on two quantum annealers,
each operating at a different temperature, we studied key mechanisms
determining their output distributions. 
In particular, we tested the freeze-out conjecture~\cite{johnson:11,Amin:2015,Amin:boltzmann}
by comparing the performance  of the two machines on
certain 
Ising problems, making use of a recent method to accurately estimate the
degeneracies of such problems.
With a working hypothesis that the output distribution is indeed a 
Boltzmann distribution of the classical problem Hamiltonian,
we calculated the effective inverse temperatures for each instance 
and machine, $\beta^{\text{eff}}$,
from which we calculated the freeze-out point.

For instances which our results show exhibit negligible 
quantum fluctuations (small $Q$), we find a well defined 
temperature-independent (i.e., machine-independent) freeze-out point, 
in agreement with the prediction of the freeze-out hypothesis 
for the small $Q$ regime.
This agreement suggests for these instances the output
distribution is indeed a classical Boltzmann distribution for $H_p$, 
with well defined effective temperature.

Our results also show, however, that for the majority of instances, 
the estimated freeze-out point is not in the regime of 
negligible quantum fluctuations, and therefore does not have a 
well-defined effective temperature, 
nor is there any reason to believe the output should follow a
classical Boltzmann distribution. 

Moreover, we also observed 
 that the effective temperatures at different programming
cycles can wildly fluctuate. Our data indicates that this
effect worsens with larger problem size. 
These observations show that for future quantum annealers to be effective
as Boltzmann samplers, designers must take into account these results, 
and find ways to ensure that instances thermalize in the
$A(s) \ll B(s)$ regime, and such that the effective temperatures are more stable.
Moving forward,  it would therefore be worthwhile to have additional estimators of
temperature, and more robust ways to reconstruct the Boltzmann distribution 
(i.e., the one we conjecture for small $Q$) \cite{global_warming}.

Promising directions include reducing sources of noise that
contribute to intrinsic control errors (ICE) in quantum annealing hardware, and
exploring alternate annealing schedules and non-standard
drivers to enable more instances
to equilibrate at a unique point late enough in the anneal that 
the quantum fluctuations are negligible.
For machine learning, another approach is possible.
It is not clear how accurately one needs to sample from Boltzmann 
distributions for machine learning, or even
that Boltzmann distributions are optimal for this purpose.
A tantalizing research  direction is the use of
distributions that have a large quantum component~\cite{Amin:boltzmann},
particularly given that certain distributions generated by quantum 
Hamiltonians are believed to have no efficient classical sampling
mechanism~\cite{q_classical_distribution,q_classical_position_momentum}. 
A deeper understanding of these processes will have profound implications for
the design of future annealers and the prospects of utilizing quantum annealers
as efficient Boltzmann samplers for machine learning and beyond.

\acknowledgments
We thank Tameem Albash, Mohammad Amin, Salvatore
Mandr{\`a} and Walter Vinci for useful discussions. 
We thank the anonymous referees for comments that significantly
improved the clarity of the manuscript.
Part of the computing
resources were provided by the USC Center for High Performance Computing and
Communications and the Oak Ridge Leadership Computing Facility at the Oak Ridge
National Laboratory, which is supported by the Office of Science of the U.S.
Department of Energy under Contract No. DE-AC05-00OR22725.  
ER would like to acknowledge support from the NASA
Advanced Exploration Systems program and NASA Ames Research Center. 
Her contributions to this work were also supported in part by the 
AFRL Information Directorate under grant F4HBKC4162G001 and the Office 
of the Director of National Intelligence (ODNI) and the Intelligence 
Advanced Research Projects Activity (IARPA), via IAA 145483. The
views and conclusions contained herein are those of the authors and should not
be interpreted as necessarily representing the official policies or
endorsements, either expressed or implied, of ODNI, IARPA, AFRL, or the U.S.
Government.  The U.S. Government is authorized to reproduce and distribute
reprints for Governmental purpose notwithstanding any copyright annotation
thereon.

\bibliography{refs.bib}

\begin{thebibliography}{35}%
\makeatletter
\providecommand \@ifxundefined [1]{%
 \@ifx{#1\undefined}
}%
\providecommand \@ifnum [1]{%
 \ifnum #1\expandafter \@firstoftwo
 \else \expandafter \@secondoftwo
 \fi
}%
\providecommand \@ifx [1]{%
 \ifx #1\expandafter \@firstoftwo
 \else \expandafter \@secondoftwo
 \fi
}%
\providecommand \natexlab [1]{#1}%
\providecommand \enquote  [1]{``#1''}%
\providecommand \bibnamefont  [1]{#1}%
\providecommand \bibfnamefont [1]{#1}%
\providecommand \citenamefont [1]{#1}%
\providecommand \href@noop [0]{\@secondoftwo}%
\providecommand \href [0]{\begingroup \@sanitize@url \@href}%
\providecommand \@href[1]{\@@startlink{#1}\@@href}%
\providecommand \@@href[1]{\endgroup#1\@@endlink}%
\providecommand \@sanitize@url [0]{\catcode `\\12\catcode `\$12\catcode
  `\&12\catcode `\#12\catcode `\^12\catcode `\_12\catcode `\%12\relax}%
\providecommand \@@startlink[1]{}%
\providecommand \@@endlink[0]{}%
\providecommand \url  [0]{\begingroup\@sanitize@url \@url }%
\providecommand \@url [1]{\endgroup\@href {#1}{\urlprefix }}%
\providecommand \urlprefix  [0]{URL }%
\providecommand \Eprint [0]{\href }%
\providecommand \doibase [0]{http://dx.doi.org/}%
\providecommand \selectlanguage [0]{\@gobble}%
\providecommand \bibinfo  [0]{\@secondoftwo}%
\providecommand \bibfield  [0]{\@secondoftwo}%
\providecommand \translation [1]{[#1]}%
\providecommand \BibitemOpen [0]{}%
\providecommand \bibitemStop [0]{}%
\providecommand \bibitemNoStop [0]{.\EOS\space}%
\providecommand \EOS [0]{\spacefactor3000\relax}%
\providecommand \BibitemShut  [1]{\csname bibitem#1\endcsname}%
\let\auto@bib@innerbib\@empty
\bibitem [{\citenamefont {Amin}(2015)}]{Amin:2015}%
  \BibitemOpen
  \bibfield  {author} {\bibinfo {author} {\bibfnamefont {M.~H.}\ \bibnamefont
  {Amin}},\ }{Searching for quantum speedup in quasistatic quantum annealers,} \href {\doibase 10.1103/PhysRevA.92.052323} {\bibfield  {journal}
  {\bibinfo  {journal} {Phys. Rev. A}\ }\textbf {\bibinfo {volume} {92}},\
  \bibinfo {pages} {052323} (\bibinfo {year} {2015})}\BibitemShut {NoStop}%
\bibitem [{\citenamefont {{M. H. Amin, E. Andriyash, J. Rolfe, B. Kulchytskyy,
  R. Melko}}(2016)}]{Amin:boltzmann}%
  \BibitemOpen
  \bibfield  {author} {\bibinfo {author} {\bibnamefont {{M. H. Amin, E.
  Andriyash, J. Rolfe, B. Kulchytskyy, R. Melko}}},\ }{Quantum Boltzmann Machine, }\href@noop {} {\bibfield
  {journal} {\bibinfo  {journal} {ArXiv e-prints}\ } (\bibinfo {year}
  {2016})},\ \Eprint {http://arxiv.org/abs/1601.02036} {arXiv:1601.02036
  [quant-ph]} \BibitemShut {NoStop}%
\bibitem [{\citenamefont {Benedetti}\ \emph {et~al.}(2016)\citenamefont
  {Benedetti}, \citenamefont {Realpe-G\'omez}, \citenamefont {Biswas},\ and\
  \citenamefont {Perdomo-Ortiz}}]{perdomo}%
  \BibitemOpen
  \bibfield  {author} {\bibinfo {author} {\bibfnamefont {M.}~\bibnamefont
  {Benedetti}}, \bibinfo {author} {\bibfnamefont {J.}~\bibnamefont
  {Realpe-G\'omez}}, \bibinfo {author} {\bibfnamefont {R.}~\bibnamefont
  {Biswas}}, \ and\ \bibinfo {author} {\bibfnamefont {A.}~\bibnamefont
  {Perdomo-Ortiz}},\ }{Estimation of effective temperatures in quantum annealers for sampling applications: A case study with possible applications in deep learning, }\href {\doibase 10.1103/PhysRevA.94.022308} {\bibfield
  {journal} {\bibinfo  {journal} {Phys. Rev. A}\ }\textbf {\bibinfo {volume}
  {94}},\ \bibinfo {pages} {022308} (\bibinfo {year} {2016})}\BibitemShut
  {NoStop}%
\bibitem [{\citenamefont {Zhang}\ \emph {et~al.}(2017)\citenamefont {Zhang},
  \citenamefont {Wagenbreth}, \citenamefont {Martin-Mayor},\ and\ \citenamefont
  {Hen}}]{fairInUnfair}%
  \BibitemOpen
  \bibfield  {author} {\bibinfo {author} {\bibfnamefont {B.~H.}\ \bibnamefont
  {Zhang}}, \bibinfo {author} {\bibfnamefont {G.}~\bibnamefont {Wagenbreth}},
  \bibinfo {author} {\bibfnamefont {V.}~\bibnamefont {Martin-Mayor}}, \ and\
  \bibinfo {author} {\bibfnamefont {I.}~\bibnamefont {Hen}},\ }{Advantages of Unfair Quantum Ground-State Sampling, }\href {\doibase
  10.1038/s41598-017-01096-6} {\bibfield  {journal} {\bibinfo  {journal}
  {Scientific Reports}\ }\textbf {\bibinfo {volume} {7}},\ \bibinfo {pages}
  {1044} (\bibinfo {year} {2017})}\BibitemShut {NoStop}%
\bibitem [{\citenamefont {Katzgraber}\ \emph {et~al.}(2015)\citenamefont
  {Katzgraber}, \citenamefont {Hamze}, \citenamefont {Zhu}, \citenamefont
  {Ochoa},\ and\ \citenamefont {Munoz-Bauza}}]{katzgraber:seekingSpeedup}%
  \BibitemOpen
  \bibfield  {author} {\bibinfo {author} {\bibfnamefont {H.~G.}\ \bibnamefont
  {Katzgraber}}, \bibinfo {author} {\bibfnamefont {F.}~\bibnamefont {Hamze}},
  \bibinfo {author} {\bibfnamefont {Z.}~\bibnamefont {Zhu}}, \bibinfo {author}
  {\bibfnamefont {A.~J.}\ \bibnamefont {Ochoa}}, \ and\ \bibinfo {author}
  {\bibfnamefont {H.}~\bibnamefont {Munoz-Bauza}},\ }{Seeking Quantum Speedup Through Spin Glasses: The Good, the Bad, and the Ugly, }\href {\doibase
  10.1103/PhysRevX.5.031026} {\bibfield  {journal} {\bibinfo  {journal} {Phys.
  Rev. X}\ }\textbf {\bibinfo {volume} {5}},\ \bibinfo {pages} {031026}
  (\bibinfo {year} {2015})}\BibitemShut {NoStop}%
\bibitem [{\citenamefont {{V. Martin-Mayor and I.
  Hen}}(2015)}]{scirep15:Martin-Mayor_Hen}%
  \BibitemOpen
  \bibfield  {author} {\bibinfo {author} {\bibnamefont {{V. Martin-Mayor and I.
  Hen}}},\ }{Unraveling Quantum Annealers using Classical Hardness, }\href {http://dx.doi.org/10.1038/srep15324} {\bibfield  {journal}
  {\bibinfo  {journal} {Scientific Reports}\ }\textbf {\bibinfo {volume} {5}},\
  \bibinfo {pages} {15324} (\bibinfo {year} {2015})}\BibitemShut {NoStop}%
\bibitem [{\citenamefont {Marshall}\ \emph {et~al.}(2016)\citenamefont
  {Marshall}, \citenamefont {Martin-Mayor},\ and\ \citenamefont
  {Hen}}]{marshall:16}%
  \BibitemOpen
  \bibfield  {author} {\bibinfo {author} {\bibfnamefont {J.}~\bibnamefont
  {Marshall}}, \bibinfo {author} {\bibfnamefont {V.}~\bibnamefont
  {Martin-Mayor}}, \ and\ \bibinfo {author} {\bibfnamefont {I.}~\bibnamefont
  {Hen}},\ }{Practical engineering of hard spin-glass instances, }\href {\doibase 10.1103/PhysRevA.94.012320} {\bibfield  {journal}
  {\bibinfo  {journal} {Phys. Rev. A}\ }\textbf {\bibinfo {volume} {94}},\
  \bibinfo {pages} {012320} (\bibinfo {year} {2016})}\BibitemShut {NoStop}%
\bibitem [{\citenamefont {{Adachi}}\ and\ \citenamefont
  {{Henderson}}(2015)}]{adachi}%
  \BibitemOpen
  \bibfield  {author} {\bibinfo {author} {\bibfnamefont {S.~H.}\ \bibnamefont
  {{Adachi}}}\ and\ \bibinfo {author} {\bibfnamefont {M.~P.}\ \bibnamefont
  {{Henderson}}},\ }{Application of Quantum Annealing to Training of Deep Neural Networks, }\href@noop {} {\bibfield  {journal} {\bibinfo  {journal}
  {ArXiv e-prints}\ } (\bibinfo {year} {2015})},\ \Eprint
  {http://arxiv.org/abs/1510.06356} {arXiv:1510.06356 [quant-ph]} \BibitemShut
  {NoStop}%
\bibitem [{\citenamefont {Johnson}\ \emph {et~al.}(2011)\citenamefont {Johnson}
  \emph {et~al.}}]{johnson:11}%
  \BibitemOpen
  \bibfield  {author} {\bibinfo {author} {\bibfnamefont {M.~W.}\ \bibnamefont
  {Johnson}} \emph {et~al.},\ }{Quantum annealing with manufactured spins, }\href {http://dx.doi.org/10.1038/nature10012}
  {\bibfield  {journal} {\bibinfo  {journal} {Nature}\ }\textbf {\bibinfo
  {volume} {473}},\ \bibinfo {pages} {194} (\bibinfo {year}
  {2011})}\BibitemShut {NoStop}%
\bibitem [{\citenamefont {King}\ and\ \citenamefont
  {McGeoch}(2014)}]{King:2014uq}%
  \BibitemOpen
  \bibfield  {author} {\bibinfo {author} {\bibfnamefont {A.~D.}\ \bibnamefont
  {King}}\ and\ \bibinfo {author} {\bibfnamefont {C.~C.}\ \bibnamefont
  {McGeoch}},\ }{Algorithm engineering for a quantum annealing platform, }\href@noop {} {\bibfield  {journal} {\bibinfo  {journal} {ArXiv
  e-prints}\ } (\bibinfo {year} {2014})},\ \Eprint
  {http://arxiv.org/abs/1410.2628} {arXiv:1410.2628 [quant-ph]} \BibitemShut
  {NoStop}%
\bibitem [{\citenamefont {Zhu}\ \emph {et~al.}(2016)\citenamefont {Zhu},
  \citenamefont {Ochoa}, \citenamefont {Schnabel}, \citenamefont {Hamze},\ and\
  \citenamefont {Katzgraber}}]{oneOverF1}%
  \BibitemOpen
  \bibfield  {author} {\bibinfo {author} {\bibfnamefont {Z.}~\bibnamefont
  {Zhu}}, \bibinfo {author} {\bibfnamefont {A.~J.}\ \bibnamefont {Ochoa}},
  \bibinfo {author} {\bibfnamefont {S.}~\bibnamefont {Schnabel}}, \bibinfo
  {author} {\bibfnamefont {F.}~\bibnamefont {Hamze}}, \ and\ \bibinfo {author}
  {\bibfnamefont {H.~G.}\ \bibnamefont {Katzgraber}},\ }{Best-case performance of quantum annealers on native spin-glass benchmarks: How chaos can affect success probabilities, }\href {\doibase
  10.1103/PhysRevA.93.012317} {\bibfield  {journal} {\bibinfo  {journal} {Phys.
  Rev. A}\ }\textbf {\bibinfo {volume} {93}},\ \bibinfo {pages} {012317}
  (\bibinfo {year} {2016})}\BibitemShut {NoStop}%
\bibitem [{\citenamefont {Boixo}\ \emph {et~al.}(2016)\citenamefont {Boixo},
  \citenamefont {Smelyanskiy}, \citenamefont {Shabani}, \citenamefont {Isakov},
  \citenamefont {Dykman}, \citenamefont {Denchev}, \citenamefont {Amin},
  \citenamefont {Smirnov}, \citenamefont {Mohseni},\ and\ \citenamefont
  {Neven}}]{oneOverF2}%
  \BibitemOpen
  \bibfield  {author} {\bibinfo {author} {\bibfnamefont {S.}~\bibnamefont
  {Boixo}}, \bibinfo {author} {\bibfnamefont {V.~N.}\ \bibnamefont
  {Smelyanskiy}}, \bibinfo {author} {\bibfnamefont {A.}~\bibnamefont
  {Shabani}}, \bibinfo {author} {\bibfnamefont {S.~V.}\ \bibnamefont {Isakov}},
  \bibinfo {author} {\bibfnamefont {M.}~\bibnamefont {Dykman}}, \bibinfo
  {author} {\bibfnamefont {V.~S.}\ \bibnamefont {Denchev}}, \bibinfo {author}
  {\bibfnamefont {M.~H.}\ \bibnamefont {Amin}}, \bibinfo {author}
  {\bibfnamefont {A.~Y.}\ \bibnamefont {Smirnov}}, \bibinfo {author}
  {\bibfnamefont {M.}~\bibnamefont {Mohseni}}, \ and\ \bibinfo {author}
  {\bibfnamefont {H.}~\bibnamefont {Neven}},\ }{Computational multiqubit tunnelling in programmable quantum annealers, }\href
  {http://dx.doi.org/10.1038/ncomms10327} {\bibfield  {journal} {\bibinfo
  {journal} {Nature Communications}\ }\textbf {\bibinfo {volume} {7}},\
  \bibinfo {pages} {10327 EP } (\bibinfo {year} {2016})}\BibitemShut {NoStop}%
\bibitem [{\citenamefont {Venturelli}\ \emph {et~al.}(2015)\citenamefont
  {Venturelli}, \citenamefont {Mandr\`a}, \citenamefont {Knysh}, \citenamefont
  {O'Gorman}, \citenamefont {Biswas},\ and\ \citenamefont
  {Smelyanskiy}}]{venturelli2015}%
  \BibitemOpen
  \bibfield  {author} {\bibinfo {author} {\bibfnamefont {D.}~\bibnamefont
  {Venturelli}}, \bibinfo {author} {\bibfnamefont {S.}~\bibnamefont
  {Mandr\`a}}, \bibinfo {author} {\bibfnamefont {S.}~\bibnamefont {Knysh}},
  \bibinfo {author} {\bibfnamefont {B.}~\bibnamefont {O'Gorman}}, \bibinfo
  {author} {\bibfnamefont {R.}~\bibnamefont {Biswas}}, \ and\ \bibinfo {author}
  {\bibfnamefont {V.}~\bibnamefont {Smelyanskiy}},\ }{Quantum Optimization of Fully Connected Spin Glasses, }\href {\doibase
  10.1103/PhysRevX.5.031040} {\bibfield  {journal} {\bibinfo  {journal} {Phys.
  Rev. X}\ }\textbf {\bibinfo {volume} {5}},\ \bibinfo {pages} {031040}
  (\bibinfo {year} {2015})}\BibitemShut {NoStop}%
\bibitem [{\citenamefont {Nifle}\ and\ \citenamefont
  {Hilhorst}(1992)}]{nifle:92}%
  \BibitemOpen
  \bibfield  {author} {\bibinfo {author} {\bibfnamefont {M.}~\bibnamefont
  {Nifle}}\ and\ \bibinfo {author} {\bibfnamefont {H.~J.}\ \bibnamefont
  {Hilhorst}},\ }{New critical-point exponent and new scaling laws for short-range Ising spin glasses, }\href {\doibase 10.1103/PhysRevLett.68.2992} {\bibfield
  {journal} {\bibinfo  {journal} {Phys. Rev. Lett.}\ }\textbf {\bibinfo
  {volume} {68}},\ \bibinfo {pages} {2992} (\bibinfo {year}
  {1992})}\BibitemShut {NoStop}%
\bibitem [{\citenamefont {Ney-Nifle}(1998)}]{ney-nifle:98}%
  \BibitemOpen
  \bibfield  {author} {\bibinfo {author} {\bibfnamefont {M.}~\bibnamefont
  {Ney-Nifle}},\ }{Chaos and universality in a four-dimensional spin glass, }\href {\doibase 10.1103/PhysRevB.57.492} {\bibfield
  {journal} {\bibinfo  {journal} {Phys. Rev. B}\ }\textbf {\bibinfo {volume}
  {57}},\ \bibinfo {pages} {492} (\bibinfo {year} {1998})}\BibitemShut
  {NoStop}%
\bibitem [{\citenamefont {Krzakala}\ and\ \citenamefont
  {Bouchaud}(2005)}]{krzakala:05}%
  \BibitemOpen
  \bibfield  {author} {\bibinfo {author} {\bibfnamefont {F.}~\bibnamefont
  {Krzakala}}\ and\ \bibinfo {author} {\bibfnamefont {J.~P.}\ \bibnamefont
  {Bouchaud}},\ }{Disorder chaos in spin glasses, }\href {\doibase 10.1209/epl/i2005-10256-2} {\bibfield
  {journal} {\bibinfo  {journal} {Europhys. Lett.}\ }\textbf {\bibinfo {volume}
  {72}},\ \bibinfo {pages} {472} (\bibinfo {year} {2005})}\BibitemShut
  {NoStop}%
\bibitem [{\citenamefont {Katzgraber}\ and\ \citenamefont
  {Krzakala}(2007)}]{katzgraber:07}%
  \BibitemOpen
  \bibfield  {author} {\bibinfo {author} {\bibfnamefont {H.~G.}\ \bibnamefont
  {Katzgraber}}\ and\ \bibinfo {author} {\bibfnamefont {F.}~\bibnamefont
  {Krzakala}},\ }{Temperature and Disorder Chaos in Three-Dimensional Ising Spin Glasses, }\href {\doibase 10.1103/PhysRevLett.98.017201} {\bibfield
  {journal} {\bibinfo  {journal} {Phys. Rev. Lett.}\ }\textbf {\bibinfo
  {volume} {98}},\ \bibinfo {pages} {017201} (\bibinfo {year}
  {2007})}\BibitemShut {NoStop}%
\bibitem [{\citenamefont {Albash}\ \emph {et~al.}(2017)\citenamefont {Albash},
  \citenamefont {Martin-Mayor},\ and\ \citenamefont {Hen}}]{tempScalingLaw}%
  \BibitemOpen
  \bibfield  {author} {\bibinfo {author} {\bibfnamefont {T.}~\bibnamefont
  {Albash}}, \bibinfo {author} {\bibfnamefont {V.}~\bibnamefont
  {Martin-Mayor}}, \ and\ \bibinfo {author} {\bibfnamefont {I.}~\bibnamefont
  {Hen}},\ }{Temperature Scaling Law for Quantum Annealing Optimizers, }\href {\doibase 10.1103/PhysRevLett.119.110502} {\bibfield
  {journal} {\bibinfo  {journal} {Phys. Rev. Lett.}\ }\textbf {\bibinfo
  {volume} {119}},\ \bibinfo {pages} {110502} (\bibinfo {year}
  {2017})}\BibitemShut {NoStop}%
\bibitem [{Note1()}]{Note1}%
  \BibitemOpen
  \bibinfo {note} {The term `effective temperature' is somewhat of a misuse as
  it may imply thermalization of the system whereas in fact it may not be the
  case.}\BibitemShut {Stop}%
\bibitem [{Note2()}]{Note2}%
  \BibitemOpen
  \bibinfo {note} {One machine is owned by Lockheed-Martin, housed at USC's
  Information Sciences Institute and the other, by a NASA-USRA-Google
  collaboration and housed inside the NASA Ames Research Center.}\BibitemShut
  {Stop}%
\bibitem [{\citenamefont {Hen}\ \emph {et~al.}(2015)\citenamefont {Hen},
  \citenamefont {Job}, \citenamefont {Albash}, \citenamefont {R\o{}nnow},
  \citenamefont {Troyer},\ and\ \citenamefont {Lidar}}]{hen:15}%
  \BibitemOpen
  \bibfield  {author} {\bibinfo {author} {\bibfnamefont {I.}~\bibnamefont
  {Hen}}, \bibinfo {author} {\bibfnamefont {J.}~\bibnamefont {Job}}, \bibinfo
  {author} {\bibfnamefont {T.}~\bibnamefont {Albash}}, \bibinfo {author}
  {\bibfnamefont {T.~F.}\ \bibnamefont {R\o{}nnow}}, \bibinfo {author}
  {\bibfnamefont {M.}~\bibnamefont {Troyer}}, \ and\ \bibinfo {author}
  {\bibfnamefont {D.~A.}\ \bibnamefont {Lidar}},\ }{Probing for quantum speedup in spin-glass problems with planted solutions, }\href {\doibase
  10.1103/PhysRevA.92.042325} {\bibfield  {journal} {\bibinfo  {journal} {Phys.
  Rev. A}\ }\textbf {\bibinfo {volume} {92}},\ \bibinfo {pages} {042325}
  (\bibinfo {year} {2015})}\BibitemShut {NoStop}%
\bibitem [{\citenamefont {Wang}\ and\ \citenamefont {Landau}(2001)}]{wang:01a}%
  \BibitemOpen
  \bibfield  {author} {\bibinfo {author} {\bibfnamefont {F.}~\bibnamefont
  {Wang}}\ and\ \bibinfo {author} {\bibfnamefont {D.~P.}\ \bibnamefont
  {Landau}},\ }{Determining the density of states for classical statistical models: A random walk algorithm to produce a flat histogram, }\href {\doibase 10.1103/PhysRevE.64.056101} {\bibfield
  {journal} {\bibinfo  {journal} {Phys. Rev. E}\ }\textbf {\bibinfo {volume}
  {64}},\ \bibinfo {pages} {056101} (\bibinfo {year} {2001})}\BibitemShut
  {NoStop}%
\bibitem [{Note3()}]{Note3}%
  \BibitemOpen
  \bibinfo {note} {If we calculate the ratio via the (least squares) gradient
  of Fig.~\ref {fig:effectiveBetas} (top), we find it is $R=1.14$, also far
  below the physical ratio.}\BibitemShut {Stop}%
\bibitem [{Note4()}]{Note4}%
  \BibitemOpen
  \bibinfo {note} {The increase in fluctuations with problem size we observe in
  Fig.~\ref {fig:betaRangeRatio} (bottom) is most likely an underestimate of
  the full effect. Since our criterion for discarding instances is convergence
  of the WL algorithm, those instances that do not appear in the figure exhibit
  fluctuations of larger magnitudes, as there is a known strong positive
  correlation between WL convergence, i.e., its classical hardness, and
  $J$-chaos (see, e.g., Ref.~\cite {scirep15:Martin-Mayor_Hen}).}\BibitemShut
  {Stop}%
\bibitem [{Note5()}]{Note5}%
  \BibitemOpen
  \bibinfo {note} {We also discount other minor effects, such as the known
  logarithmic dependence on anneal time, in Appendix \ref
  {sect:additional}.}\BibitemShut {Stop}%
\bibitem [{\citenamefont {Raymond}\ \emph {et~al.}(2016)\citenamefont
  {Raymond}, \citenamefont {Yarkoni},\ and\ \citenamefont
  {Andriyash}}]{global_warming}%
  \BibitemOpen
  \bibfield  {author} {\bibinfo {author} {\bibfnamefont {J.}~\bibnamefont
  {Raymond}}, \bibinfo {author} {\bibfnamefont {S.}~\bibnamefont {Yarkoni}}, \
  and\ \bibinfo {author} {\bibfnamefont {E.}~\bibnamefont {Andriyash}},\ }{Global Warming: Temperature Estimation in Annealers, }\href
  {\doibase 10.3389/fict.2016.00023} {\bibfield  {journal} {\bibinfo  {journal}
  {Frontiers in ICT}\ }\textbf {\bibinfo {volume} {3}},\ \bibinfo {pages} {23}
  (\bibinfo {year} {2016})}\BibitemShut {NoStop}%
\bibitem [{\citenamefont {Semay}\ and\ \citenamefont
  {Ducobu}(2016)}]{q_classical_distribution}%
  \BibitemOpen
  \bibfield  {author} {\bibinfo {author} {\bibfnamefont {C.}~\bibnamefont
  {Semay}}\ and\ \bibinfo {author} {\bibfnamefont {L.}~\bibnamefont {Ducobu}},\
  }{Quantum and classical probability distributions for arbitrary Hamiltonians, }\href {http://stacks.iop.org/0143-0807/37/i=4/a=045403} {\bibfield
  {journal} {\bibinfo  {journal} {European Journal of Physics}\ }\textbf
  {\bibinfo {volume} {37}},\ \bibinfo {pages} {045403} (\bibinfo {year}
  {2016})}\BibitemShut {NoStop}%
\bibitem [{\citenamefont {{R. W.
  Robinett}}(1995)}]{q_classical_position_momentum}%
  \BibitemOpen
  \bibfield  {author} {\bibinfo {author} {\bibnamefont {{R. W. Robinett}}},\
  }{Quantum and classical probability distributions for position and momentum, }\href {http://dx.doi.org/10.1119/1.17807} {\bibfield  {journal} {\bibinfo
  {journal} {{Am. J. Phys.}}\ }\textbf {\bibinfo {volume} {63}},\ \bibinfo
  {pages} {823} (\bibinfo {year} {1995})}\BibitemShut {NoStop}%
\bibitem [{\citenamefont {Bunyk}\ \emph {et~al.}(2014)\citenamefont {Bunyk},
  \citenamefont {Hoskinson}, \citenamefont {Johnson}, \citenamefont
  {Tolkacheva}, \citenamefont {Altomare}, \citenamefont {Berkley},
  \citenamefont {Harris}, \citenamefont {Hilton}, \citenamefont {Lanting},
  \citenamefont {Przybysz},\ and\ \citenamefont {Whittaker}}]{Bunyk:2014hb}%
  \BibitemOpen
  \bibfield  {author} {\bibinfo {author} {\bibfnamefont {P.~I.}\ \bibnamefont
  {Bunyk}}, \bibinfo {author} {\bibfnamefont {E.~M.}\ \bibnamefont
  {Hoskinson}}, \bibinfo {author} {\bibfnamefont {M.~W.}\ \bibnamefont
  {Johnson}}, \bibinfo {author} {\bibfnamefont {E.}~\bibnamefont {Tolkacheva}},
  \bibinfo {author} {\bibfnamefont {F.}~\bibnamefont {Altomare}}, \bibinfo
  {author} {\bibfnamefont {A.}~\bibnamefont {Berkley}}, \bibinfo {author}
  {\bibfnamefont {R.}~\bibnamefont {Harris}}, \bibinfo {author} {\bibfnamefont
  {J.~P.}\ \bibnamefont {Hilton}}, \bibinfo {author} {\bibfnamefont
  {T.}~\bibnamefont {Lanting}}, \bibinfo {author} {\bibfnamefont
  {A.}~\bibnamefont {Przybysz}}, \ and\ \bibinfo {author} {\bibfnamefont
  {J.}~\bibnamefont {Whittaker}},\ }{Architectural Considerations in the Design of a Superconducting Quantum Annealing Processor, }\href
  {\doibase 10.1109/TASC.2014.2318294} {\bibfield  {journal} {\bibinfo
  {journal} {Applied Superconductivity, IEEE Transactions on}\ }\textbf
  {\bibinfo {volume} {24}},\ \bibinfo {pages} {1} (\bibinfo {year} {Aug.
  2014})}\BibitemShut {NoStop}%
\bibitem [{\citenamefont {Choi}(2011)}]{Choi2}%
  \BibitemOpen
  \bibfield  {author} {\bibinfo {author} {\bibfnamefont {V.}~\bibnamefont
  {Choi}},\ }{Minor-embedding in adiabatic quantum computation: II. Minor-universal graph design, }\href {\doibase 10.1007/s11128-010-0200-3} {\bibfield  {journal}
  {\bibinfo  {journal} {Quant. Inf. Proc.}\ }\textbf {\bibinfo {volume} {10}},\
  \bibinfo {pages} {343} (\bibinfo {year} {2011})}\BibitemShut {NoStop}%
\bibitem [{\citenamefont {Barthel}\ \emph {et~al.}(2002)\citenamefont
  {Barthel}, \citenamefont {Hartmann}, \citenamefont {Leone}, \citenamefont
  {Ricci-Tersenghi}, \citenamefont {Weigt},\ and\ \citenamefont
  {Zecchina}}]{Barthel:2002tw}%
  \BibitemOpen
  \bibfield  {author} {\bibinfo {author} {\bibfnamefont {W.}~\bibnamefont
  {Barthel}}, \bibinfo {author} {\bibfnamefont {A.~K.}\ \bibnamefont
  {Hartmann}}, \bibinfo {author} {\bibfnamefont {M.}~\bibnamefont {Leone}},
  \bibinfo {author} {\bibfnamefont {F.}~\bibnamefont {Ricci-Tersenghi}},
  \bibinfo {author} {\bibfnamefont {M.}~\bibnamefont {Weigt}}, \ and\ \bibinfo
  {author} {\bibfnamefont {R.}~\bibnamefont {Zecchina}},\ }{Hiding Solutions in Random Satisfiability Problems: A Statistical Mechanics Approach, }\href
  {http://link.aps.org/doi/10.1103/PhysRevLett.88.188701} {\bibfield  {journal}
  {\bibinfo  {journal} {Phys. Rev. Lett.}\ }\textbf {\bibinfo {volume} {88}},\
  \bibinfo {pages} {188701} (\bibinfo {year} {2002})}\BibitemShut {NoStop}%
\bibitem [{\citenamefont {Krzakala}\ and\ \citenamefont
  {Zdeborov{\'a}}(2009)}]{Krzakala:2009qo}%
  \BibitemOpen
  \bibfield  {author} {\bibinfo {author} {\bibfnamefont {F.}~\bibnamefont
  {Krzakala}}\ and\ \bibinfo {author} {\bibfnamefont {L.}~\bibnamefont
  {Zdeborov{\'a}}},\ }{Hiding Quiet Solutions in Random Constraint Satisfaction Problems, }\href
  {http://link.aps.org/doi/10.1103/PhysRevLett.102.238701} {\bibfield
  {journal} {\bibinfo  {journal} {Phys. Rev. Lett.}\ }\textbf {\bibinfo
  {volume} {102}},\ \bibinfo {pages} {238701} (\bibinfo {year}
  {2009})}\BibitemShut {NoStop}%
\bibitem [{\citenamefont {Boixo}\ \emph {et~al.}(2014)\citenamefont {Boixo},
  \citenamefont {R{\o}nnow}, \citenamefont {Isakov}, \citenamefont {Wang},
  \citenamefont {Wecker}, \citenamefont {Lidar}, \citenamefont {Martinis},\
  and\ \citenamefont {Troyer}}]{q108}%
  \BibitemOpen
  \bibfield  {author} {\bibinfo {author} {\bibfnamefont {S.}~\bibnamefont
  {Boixo}}, \bibinfo {author} {\bibfnamefont {T.~F.}\ \bibnamefont
  {R{\o}nnow}}, \bibinfo {author} {\bibfnamefont {S.~V.}\ \bibnamefont
  {Isakov}}, \bibinfo {author} {\bibfnamefont {Z.}~\bibnamefont {Wang}},
  \bibinfo {author} {\bibfnamefont {D.}~\bibnamefont {Wecker}}, \bibinfo
  {author} {\bibfnamefont {D.~A.}\ \bibnamefont {Lidar}}, \bibinfo {author}
  {\bibfnamefont {J.~M.}\ \bibnamefont {Martinis}}, \ and\ \bibinfo {author}
  {\bibfnamefont {M.}~\bibnamefont {Troyer}},\ }{Evidence for quantum annealing with more than one hundred qubits, }\href {\doibase
  10.1038/nphys2900} {\bibfield  {journal} {\bibinfo  {journal} {Nat. Phys.}\
  }\textbf {\bibinfo {volume} {10}},\ \bibinfo {pages} {218} (\bibinfo {year}
  {2014})}\BibitemShut {NoStop}%
\bibitem [{\citenamefont {R{\o}nnow}\ \emph {et~al.}(2014)\citenamefont
  {R{\o}nnow}, \citenamefont {Wang}, \citenamefont {Job}, \citenamefont
  {Boixo}, \citenamefont {Isakov}, \citenamefont {Wecker}, \citenamefont
  {Martinis}, \citenamefont {Lidar},\ and\ \citenamefont {Troyer}}]{speedup}%
  \BibitemOpen
  \bibfield  {author} {\bibinfo {author} {\bibfnamefont {T.~F.}\ \bibnamefont
  {R{\o}nnow}}, \bibinfo {author} {\bibfnamefont {Z.}~\bibnamefont {Wang}},
  \bibinfo {author} {\bibfnamefont {J.}~\bibnamefont {Job}}, \bibinfo {author}
  {\bibfnamefont {S.}~\bibnamefont {Boixo}}, \bibinfo {author} {\bibfnamefont
  {S.~V.}\ \bibnamefont {Isakov}}, \bibinfo {author} {\bibfnamefont
  {D.}~\bibnamefont {Wecker}}, \bibinfo {author} {\bibfnamefont {J.~M.}\
  \bibnamefont {Martinis}}, \bibinfo {author} {\bibfnamefont {D.~A.}\
  \bibnamefont {Lidar}}, \ and\ \bibinfo {author} {\bibfnamefont
  {M.}~\bibnamefont {Troyer}},\ }{Defining and detecting quantum speedup, }\href {\doibase 10.1126/science.1252319}
  {\bibfield  {journal} {\bibinfo  {journal} {Science}\ }\textbf {\bibinfo
  {volume} {345}},\ \bibinfo {pages} {420} (\bibinfo {year}
  {2014})}\BibitemShut {NoStop}%
\bibitem [{\citenamefont {{M. Mezard, G. Parisi and M.A.
  Virasoro}}(1987)}]{Parisi:book}%
  \BibitemOpen
  \bibfield  {author} {\bibinfo {author} {\bibnamefont {{M. Mezard, G. Parisi
  and M.A. Virasoro}}},\ }\href@noop {} {\emph {\bibinfo {title} {{Spin Glass
  Theory and Beyond}}}},\ {World Scientific Lecture Notes in Physics}\
  (\bibinfo  {publisher} {{World Scientific}},\ \bibinfo {address}
  {{Singapore}},\ \bibinfo {year} {1987})\BibitemShut {NoStop}%
\end{thebibliography}%

\appendix

\section{The Google-NASA-USRA (`cold') and Lockheed-Martin-USC (`hot') D-Wave Two processors\label{sect:devices}}
 
The quantum annealer used in our work is the D-Wave Two (DW2) device \cite{Bunyk:2014hb}. This device is designed to solve optimization problems by evolving a known initial configuration --- the ground state of a transverse field $H_d =-\sum_i \sigma_i^x$, where $\sigma_i^x$ is the Pauli spin-$1/2$ matrix acting on spin $i$ --- towards the ground state of a classical Ising-model Hamiltonian which serves as a cost function that encodes the problem that is to be solved:
\beq 
H_p =\sum_{\langle i,j\rangle} J_{ij} \sigma_i^z \sigma_j^z + \sum_i h_i \sigma_i^z \ .
\label{eq:H}
\eeq 
The variables $\{\sigma_i^z\}$ denote either classical Ising-spin variables that take values $\pm1$ or Pauli spin-$1/2$ matrices, the $\{J_{ij}\}$ are programmable coupling parameters, and the $\{h_{i}\}$ are programmable local longitudinal fields. The $N$ spin variables are realized as superconducting flux qubits and occupy the vertices of the D-Wave `Chimera' hardware graph~\cite{Bunyk:2014hb,Choi2}. Here, $
\langle i,j\rangle$ denotes summation over the edges of the graph. The union of the two D-Wave Chimera graphs is given in Fig.~\ref{fig:chimera} -- this is the graph all of our problem instances were defined on.

\begin{figure}[htp]
\begin{center}
\includegraphics[width=0.9\columnwidth]{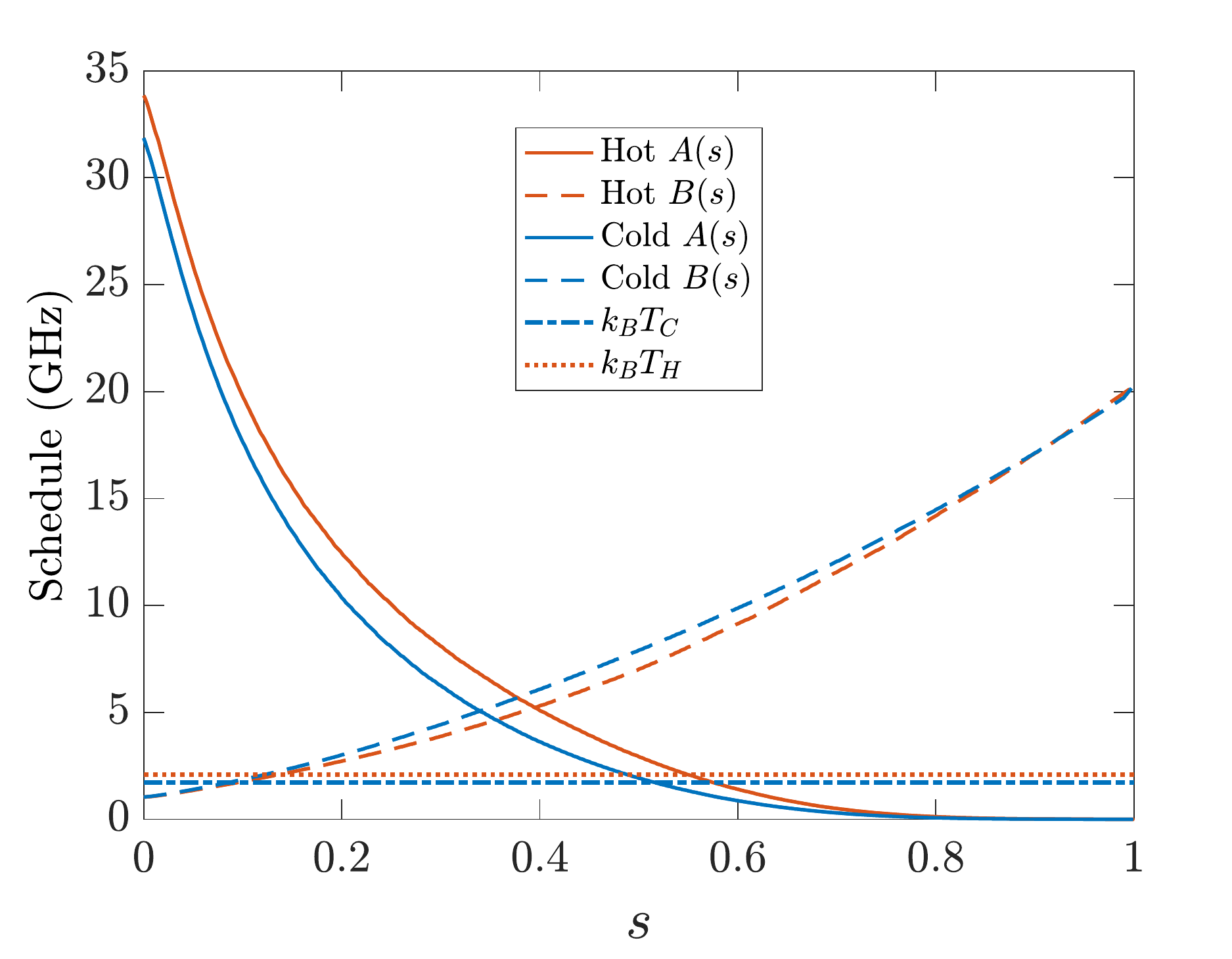}
\caption{\textbf{Annealing schedules of the USC (hot) and NASA (cold) DW2 processors.}
Annealing schedule [see Eq.~(\ref{annealing2})] in GHz as a function of
dimensionless annealing time $s=t/\mathcal{T}$. We also plot the temperatures
($\hbar=1$) of the devices (see legend).}
\label{fig:schedule2}
\end{center}
\end{figure}

These machines evolve the full Hamiltonian via 
\beq\label{annealing2}
H(s)=A(s)H_d+B(s)H_p \,.
\eeq
The way in which the strength of the initial ($H_d$) and final ($H_p$) Hamiltonians evolve is given by the parameters $A(s)$ and $B(s)$, where $s=t/\mathcal{T} \in [0,1]$ is the annealing time. Here, $\mathcal{T}$ is the total annealing time, ranging between $20 \mu$s and $20 m$s on these devices. The annealing schedule is given in Fig.~\ref{fig:schedule2} for each machine.

\begin{figure}[htp]
\begin{center}
\includegraphics[width=0.9\columnwidth]{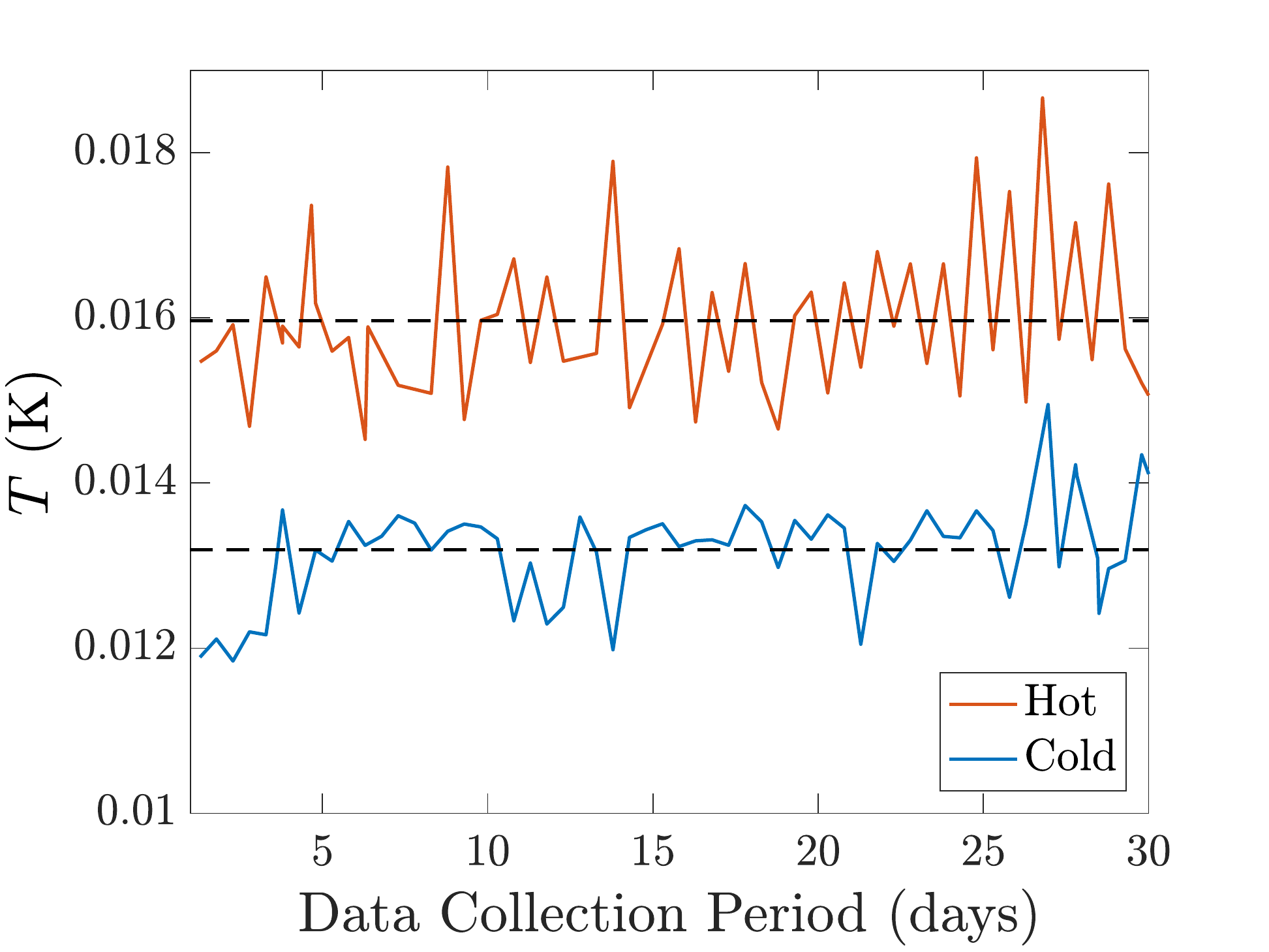}
\caption{\textbf{Temperature logs}. Temperature log for the USC (hot) and NASA
(cold) machines during the period of time we performed our experiments. 
The dashed black lines represent the mean of the data sets. 
The mean temperature of the `hot' USC machine was about $T_H = 16.0$ mK, 
and the mean temperature of the `cold' NASA machine was about $T_C = 13.2$ mK, 
with ratio $T_H/T_C \approx 1.21$. 
Note, the temperature data is sparse, sampled only twice per
24 hours.} \label{fig:tempLogs}
\end{center}
\end{figure}

In Fig.~\ref{fig:tempLogs} we show the temperature log of the D-Wave chips
during the time which we collected our data.  

\section{Generation of instances\label{sec:instances}}
For the generation of instances, we have chosen in this work to study problems constructed around `planted solutions' -- an idea borrowed from constraint satisfaction (SAT) problems~\cite{Barthel:2002tw,Krzakala:2009qo}. In these problems, the planted solution represents a ground state configuration of Eq.~\eqref{eq:H} that minimizes the energy and is known in advance. This knowledge circumvents the need to verify the ground state energy using exact (provable) solvers, which rapidly become too expensive computationally as the number of variables grows, and which were employed in earlier benchmarking studies \cite{q108,speedup}. Moreover, these problems are known to possess different degrees of `tunable hardness', achieved by adjusting the amount of frustration (see Ref.~\cite{Parisi:book}) which we will use. Last, studying this type of problems allows us to devise an algorithm to find all minimizing configurations of the generated instances. Knowledge of both energy levels as their degeneracies is essential for the calculation of effective temperatures of the instances. 

To generate the instances for the experiment we follow the guidelines introduced in Ref.~\cite{hen:15}. We generate 13 groups of 100 instances, for each of 7 different sub-Chimera sizes with
$L=2\dots8$, see Fig.~\ref{fig:chimera} (i.e.,  9100 total instances). These 13
groups differ in the ratio $\alpha$ of number of clauses (or loops) to number of qubits contained in each instance.
For every fixed ratio $\alpha$ the range of integer-valued $J_{ij}$'s, which we denote by $J_{\max}$ is fixed across the different problem sizes. 
D-Wave further rescales all coupling values such that the encoded values, $\tilde{J}\in [-1,1]$ which implies that for every fixed $\alpha$ both the range of $J$-values as well as the spacings between them is identical across different problem sizes.

\section{Wang-Landau entropic sampler\label{sect:WL}}
As explained in the main text, we employed a Wang-Landau entropic sampler to estimate the degeneracy of the energy levels for our generated planted-solution instances. This algorithm performs essentially a random walk over the energy landscape, where updates at each step in the algorithm are such that an approximately flat histogram of visited energies is produced. We follow the same methodology as originally described in \cite{wang:01a}. Our histogram was considered `flat' when the lowest sampled energy level has been visited at least 80\% of the mean of the entire histogram.

We performed 20 independent Wang-Landau runs, each up to $10^9$ steps for each of our instances. We then averaged over these 20 runs which provided our estimate of degeneracies for each instance.
We then discarded any instances for which the ground or first excited state
degeneracies did not match that for the exact solution counter (up to 5\%
error). This meant we had accurate degeneracy data for problems up to 282
qubits in size.

\section{Effects or lack thereof of additional experimental parameters\label{sect:additional}}

\subsection{Success probabilities}

\begin{figure}[htp]
\begin{center}
\includegraphics[scale=0.55]{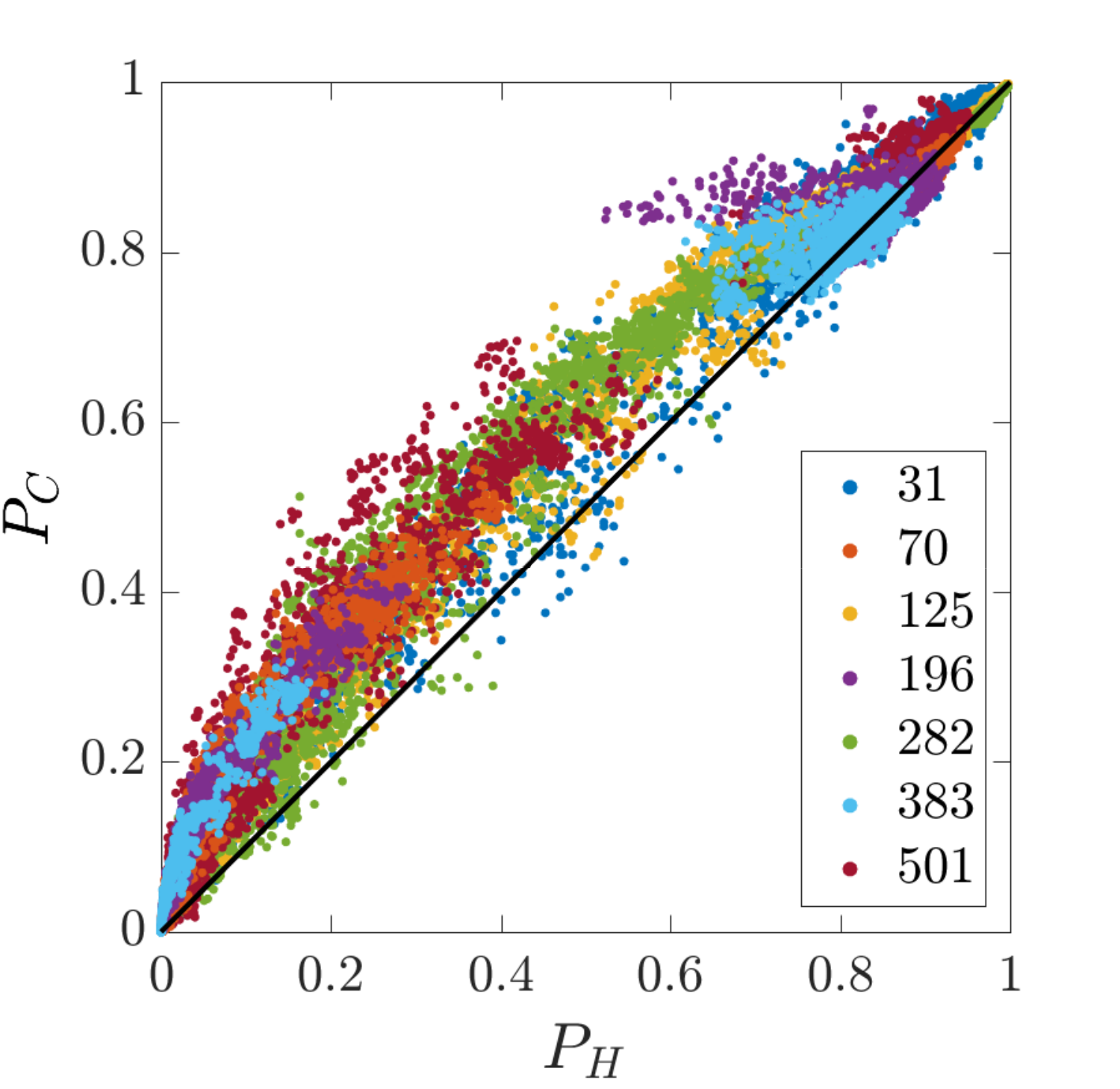}
\caption{\textbf{Success probability}. Probability of success (how often the ground state energy is correctly identified) of all instances and programming cycles on the two machines (`hot' USC machine $P_H$, and the `cold' NASA machine $P_C$). Each point is a randomly chosen programming cycle (for the same instance on each machine). Number of qubits given by legend.}
\label{fig:Psuccess}
\end{center}
\end{figure}

In Fig.~\ref{fig:Psuccess} we show the histogram of the success probabilities for the two machines, for all of the instances. We see the cold (NASA) machine clearly outperforms the hot (USC) machine--we expect, due to the colder operating temperature.

\subsection{Programming cycles}

\begin{figure}[htp]
\begin{center}
\includegraphics[width=\columnwidth]{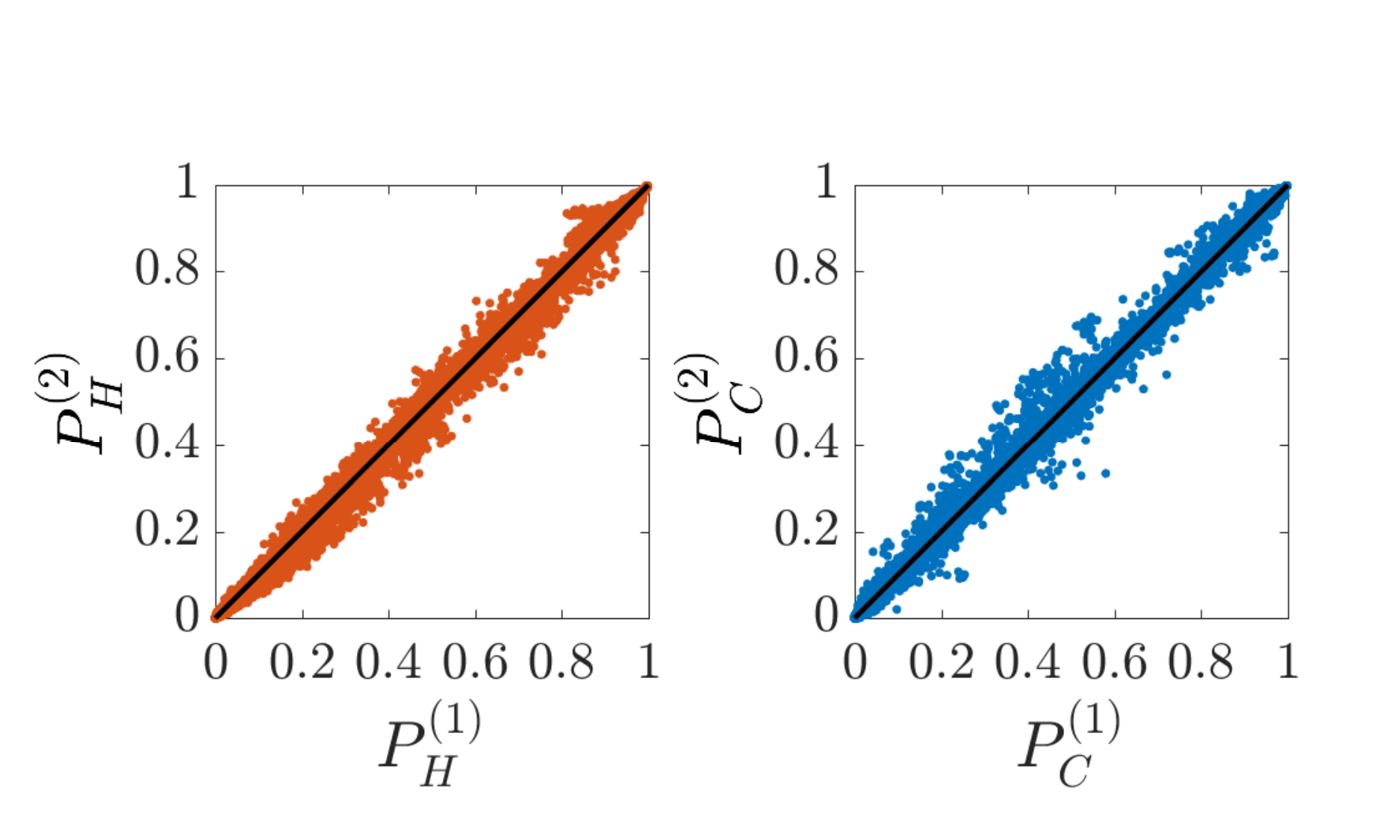}
\caption{\textbf{Machine correlation}. We compare the results of two programming cycles for each instance on each machine. We see the data aligns nicely along $y=x$, albeit with sizable fluctuations (as one would expect). Compare this with Fig.~\ref{fig:Psuccess}, where data clearly deviates from $y=x$. The `hot' USC machine is on the left, and the `cold' NASA machine on the right.}
\label{fig:machineCorrelation}
\end{center}
\end{figure}

In Fig.~\ref{fig:machineCorrelation} we compare two different programming cycles (from different days) on the same machine, showing consistency over different runs. 

\subsection{Anneal times}
\begin{figure}[htp]
\begin{center}
\includegraphics[width=\columnwidth]{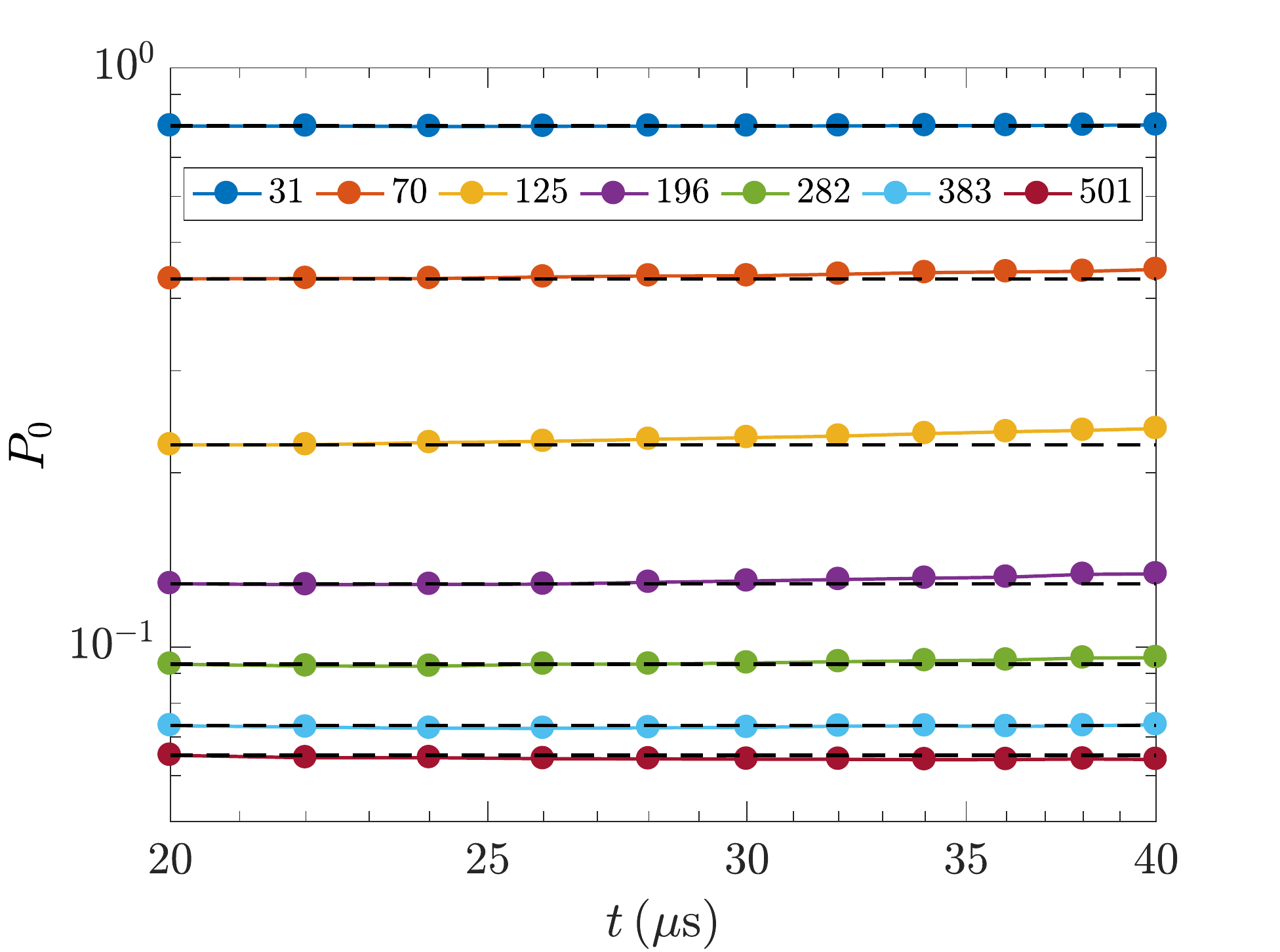}
\caption{\textbf{D-Wave success with anneal time}. Average probability of success, $P_0$ (for the hotter USC machine), against anneal time (log scale), for different problem sizes (see legend). Each point averaged over two programming cycles, of $N_{\text{anneals}}=20,000$ anneals each. The black dash lines correspond to the value of $P_0$ for $t=20$ $\mu$s. One can see in general a slight increase in success probability with $t$.}
\label{fig:annealTime}
\end{center}
\end{figure}

\begin{figure}[htp]
\begin{center}
\includegraphics[width=0.9\columnwidth]{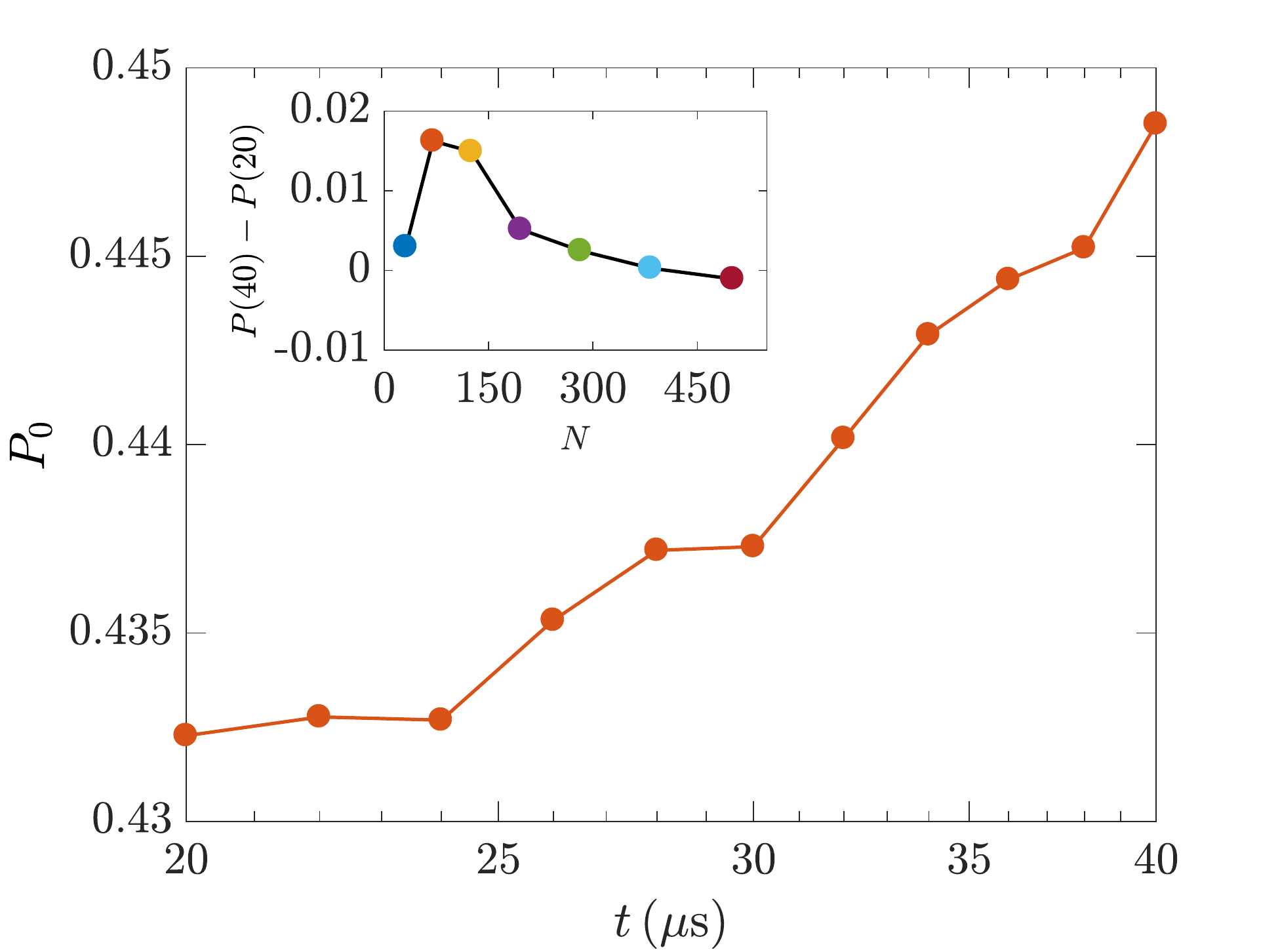}
\caption{\textbf{D-Wave success with anneal time, for single problem size}. Average probability of success, $P_0$ (for the hotter USC machine), against anneal time (log scale), for problem size 70 ($L=3$). Note the approximate linear relationship. Inset: Difference in success probability between $t=40$ and $t=20$ $\mu$s as measured by $P(40)-P(20)$, where $P(t)$ is defined as $P_0$ for anneal time $t$.}
\label{fig:annealTime70}
\end{center}
\end{figure}

We also study the effect of varying anneal time on success probability in Figs.~\ref{fig:annealTime} and \ref{fig:annealTime70}. We see that there is only a very weak (logarithmic) dependence on anneal time, in accordance with \cite{Amin:2015,scirep15:Martin-Mayor_Hen}, and moreover, it is seemingly not correlated with problem size.

\subsection{Ratio of number of clauses to number of qubits}
\begin{figure}[htp]
\begin{center}
\includegraphics[width=0.9\columnwidth]{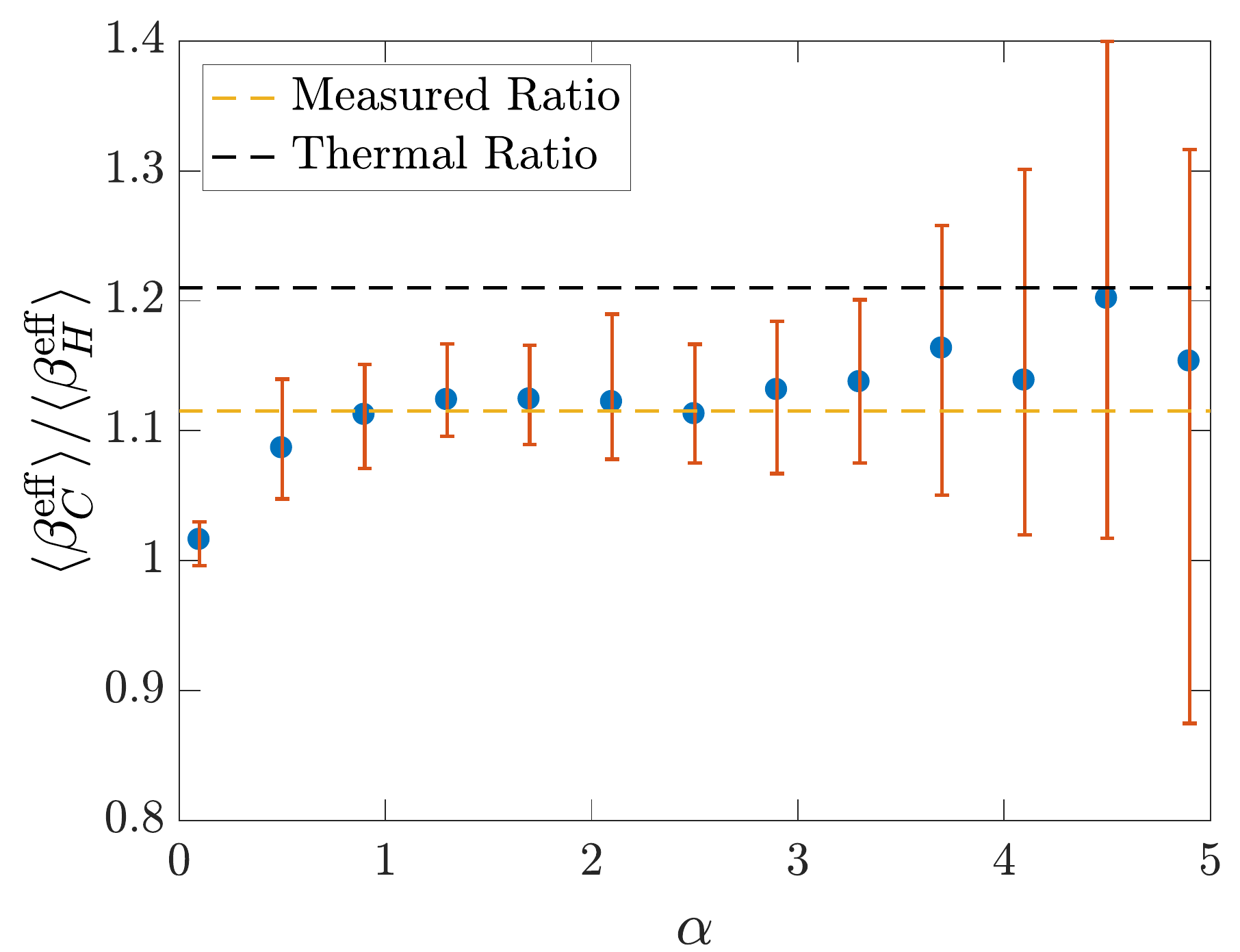}
\caption{\textbf{Effective inverse temperature ratio against clause density}. As explained in the text, we generated many problems for each graph size $L=2,\dots,8$. We picked fixed and identical values of $\alpha$ (the ratio of the number of loops to the number of qubits) for each qubit size. We find that $\alpha$ has (within the 95\% confidence interval error bars obtained from bootstrapping the data) mostly no effect on the effective temperature ratio we measured from our data (1.11, calculated as the median over all instances), and that it is much below the thermal ratio value of 1.21. Note the smallest $\alpha=0.1$ instances are very easy to solve ($P_0 \approx 1$), making it hard to distinguish performance (hence temperature) differences between the machines.}
\label{fig:ratio_alpha}
\end{center}
\end{figure}

Fig.~\ref{fig:ratio_alpha} shows that (within error bars) our computation of the inverse temperature ratio between the two machines is unaffected by changing the ratio of number of clauses (or loops) to number of qubits $\alpha$, and moreover, that the thermal ratio is well above the measured ratio (far outside of the error bars mostly).

\begin{figure}[H]
\begin{center}
\includegraphics[width=0.9\columnwidth]{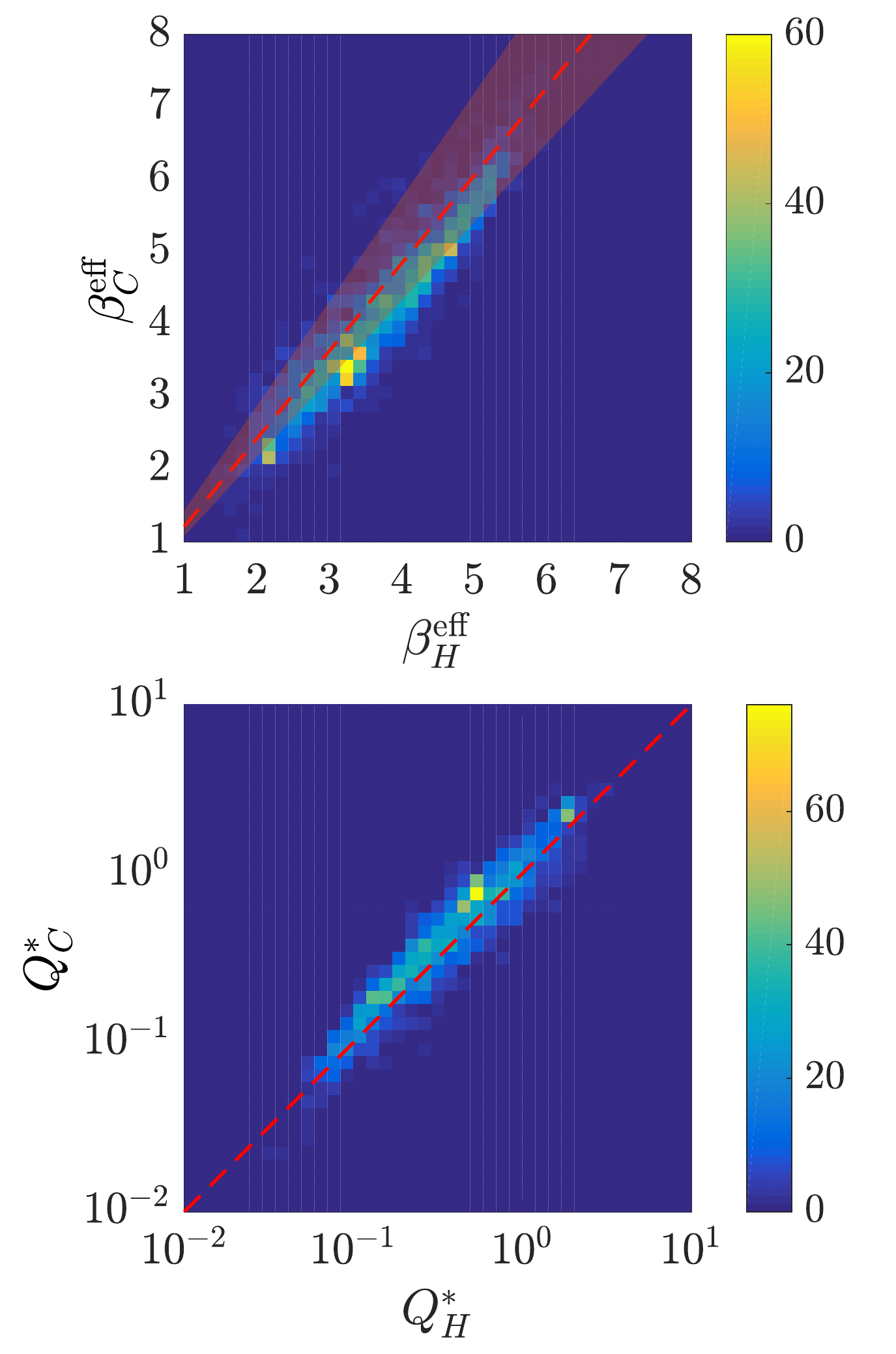}
\caption{\textbf{Density plots for $\beta^{\text{eff}}$ and $Q^*$}. The corresponding `heat map' for Fig. 2 of the main text. Color indicates the number of points in each region (given by color bar on the right hand side). Top: Red dash line is the `ideal' thermal ratio (i.e. if the ratio of the effective inverse temperatures were the same as the physical `thermal' inverse temperatures). The variance in physical temperature fluctuations is given by the red semi-transparent region. Bottom: Red dash line is $y=x$.}
\label{fig:heat}
\end{center}
\end{figure}

\section{Additional data}
 
 Relating to Fig.~\ref{fig:effectiveBetas} of the main text, we produce a `heat map', Fig.~\ref{fig:heat}, for $\beta^{\text{eff}}$, and for $Q^*$ (defined in main text), for each machine, which shows the number of instances found in each small region. We notice that in the upper figure that the larger the effective inverse temperature, more instances fall within the `thermal region', indicating a stronger dependence on the temperature for these instances. These instances correspond to the ones which freezeout at a later point in the anneal, and thus a smaller $Q^*$ in the lower figure. In this lower figure, we observe the fit is closer to the `ideal' $y=x$ for smaller $Q^*$, and it deviates above this line for larger $Q^*$ (we discuss in more detail in the main text).

\begin{figure}[H]
\begin{center}
\includegraphics[width=0.9\columnwidth]{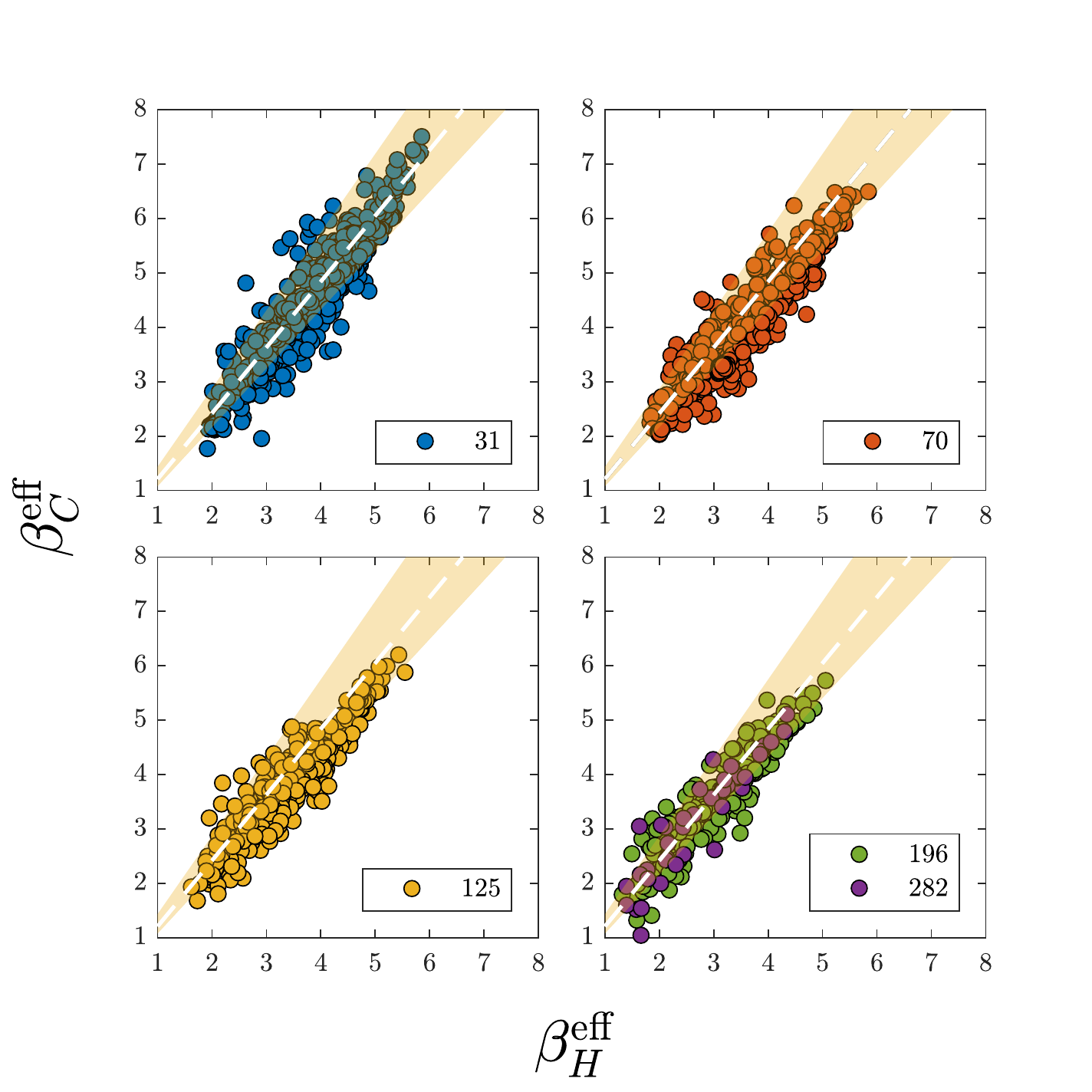}
\caption{\textbf{$\beta^{\text{eff}}$ for different problem sizes}. These plots show the effective inverse temperature found by each machine (and for each instance), for the 5 different problem sizes (see legend) for which we have reliable degeneracy data. White dash line is the `ideal' thermal ratio (i.e. if the ratio of the effective inverse temperatures were the same as the physical `thermal' inverse temperatures).  The variance in physical temperature fluctuations is given by the yellow semi-transparent region.}
\label{fig:betaNq}
\end{center}
\end{figure}

\begin{figure}[H]
\begin{center}
\includegraphics[width=0.9\columnwidth]{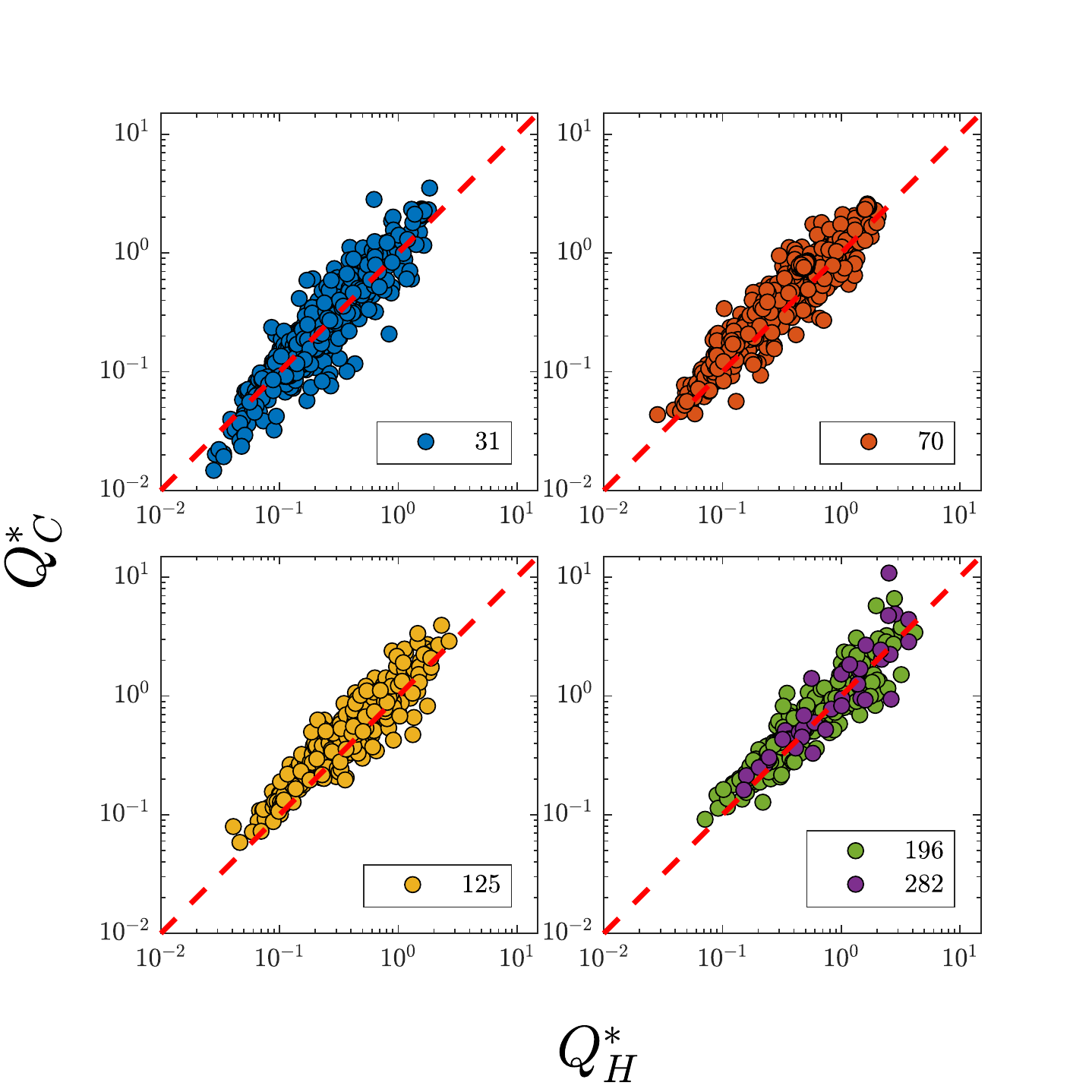}
\caption{\textbf{$Q^*$ for different problem sizes}. These plots show the median $Q^*$ found by each machine (and for each instance), for the 5 different problem sizes (see legend) for which we have reliable degeneracy data. Red dash line is $y=x$.}
\label{fig:Qnq}
\end{center}
\end{figure}

Also relating to Fig.~\ref{fig:effectiveBetas} of the main text, we produce Figs.~\ref{fig:betaNq}, \ref{fig:Qnq}, which plot $\beta^{\text{eff}}$, and $Q^*$, for each machine, but split up by problem size. One can see that typically the larger problems  exhibit lower values of $\beta^{\text{eff}}$, and likewise, larger values of $Q^*$, indicating these are in fact not thermalizing according to a Boltzmann distribution.

\newpage

\end{document}